\documentclass[twoside]{scrartcl}
\usepackage[T1]{fontenc}
\usepackage[latin9]{inputenc}
\usepackage{units}
\usepackage{array}
\usepackage{graphicx}
\usepackage{epstopdf}
\usepackage{float}
\usepackage{amsmath}
\usepackage{amssymb}
\usepackage{esint}
\usepackage{multicol}
\usepackage{changepage}

\usepackage{xcolor}
\usepackage{float}
\usepackage{placeins}

\usepackage[justification=justified]{caption}
\usepackage{prettyref}
\newrefformat{sfig}{Figure \ref{#1}}
\newrefformat{tab}{Supp. Table \ref{#1}}
\DeclareCaptionLabelFormat{supptable}{#1 S#2} 
\DeclareCaptionLabelFormat{suppfig}{#1 S#2} 

\newcommand{\showfigures}[1]{{#1}} 

\newcommand{\dx}{\ensuremath{l}}
\newcommand{\sig}{\ensuremath{\sigma}}
\newcommand{\sxx}{\ensuremath{\sigma_{xx}}}
\newcommand{\syy}{\ensuremath{\sigma_{yy}}}
\newcommand{\sxy}{\ensuremath{\sigma_{xy}}}
\newcommand{\syx}{\ensuremath{\sigma_{yx}}}

\newcommand{\sigxx}[2]{\ensuremath{\sigma_{xx}(#1,#2)}}
\newcommand{\sigyy}[2]{\ensuremath{\sigma_{yy}(#1,#2)}}
\newcommand{\sigxy}[2]{\ensuremath{\sigma_{xy}(#1,#2)}}
\newcommand{\sigyx}[2]{\ensuremath{\sigma_{yx}(#1,#2)}}
\newcommand{\vT}[2]{\ensuremath{{\vec t}(#1,#2)}}
\newcommand{\Txij}[2]{\ensuremath{t_x(#1,#2)}}
\newcommand{\Tyij}[2]{\ensuremath{t_y(#1,#2)}}
\newcommand{\srr}{\ensuremath{\sigma_{rr}}}
\newcommand{\stt}{\ensuremath{\sigma_{\theta\theta}}}
\newcommand{\srt}{\ensuremath{\sigma_{r\theta}}}
\newcommand{\str}{\ensuremath{\sigma_{\theta r}}}
\newcommand{\Tx}{\ensuremath{t_{x}}}
\newcommand{\Ty}{\ensuremath{t_{y}}}

\newcommand{\Tr}{\ensuremath{t_{r}}}
\newcommand{\Tt}{\ensuremath{t_{\theta}}}
\newcommand{\DP}[2]{\ensuremath{\frac{\partial{#1}}{\partial{#2}}}}
\newcommand{\axy}{\ensuremath{\alpha_{xy}}}
\newcommand{\aBC}{\ensuremath{\alpha_{\mathrm{BC}}}}



\begin{document}

\begin{flushleft}
{\Large
\textbf{Inference of Internal Stress in a Cell Monolayer}
}
\newline
\\
V.~Nier\textsuperscript{1},
S.~Jain\textsuperscript{2},
C. T.~Lim\textsuperscript{2,3},
S.~Ishihara\textsuperscript{4},
B.~Ladoux\textsuperscript{2,5} and
P.~Marcq\textsuperscript{1,*}
\\
\bigskip
\bf{1} Sorbonne Universit\'es, UPMC Univ Paris 6, 
Institut Curie, CNRS, UMR 168, Laboratoire Physco-Chimie Curie, Paris, France
\\
\bf{2} Mechanobiology Institute, National University of Singapore,
Singapore
\\
\bf{3} Department of Biomedical Engineering and Department of
Mechanical Engineering, National University of Singapore, Singapore 
\\
\bf{4} Department of Physics, Meiji University, Kawasaki, 
Kanagawa, Japan
\\
\bf{5} Institut Jacques Monod, CNRS UMR 7592 and Universit\'e Paris
Diderot, Paris, France
\\
\bigskip
* Corresponding Author philippe.marcq@curie.fr
\end{flushleft}

\date{\today}

\section*{Abstract}

We combine  traction force data with Bayesian inversion to obtain an absolute
estimate of the internal stress field of a cell monolayer.
The method, Bayesian inversion stress microscopy (BISM), is validated 
using numerical simulations performed in a wide range of conditions.
It is robust to changes in each ingredient of the underlying statistical model.
Importantly, its accuracy does not depend on the rheology
of the tissue. We apply BISM to experimental traction force data
measured in a narrow ring of cohesive epithelial cells, and check that
the inferred stress field coincides with that obtained by direct
spatial integration of the traction force data in this 
quasi-one-dimensional geometry.

\section{Introduction}
\label{sec:Introduction}

Dynamical behaviors of multicellular assemblies play a crucial role during 
tissue development \cite{Keller2000} and in the maintenance of adult tissues
\cite{vanderFlier2009}. In addition, disregulation of multicellular
structures may lead to pathological situations such as tumor
formation and tumor progression \cite{Friedl2009}. In this context, 
cell monolayers have been extensively studied to model \emph{in vivo} 
tissue functions. Such approaches allow for well-controlled experiments, 
which have been performed in a variety of settings, such as
monolayer spreading \cite{Poujade2007,Trepat2009},
wound healing \cite{Anon2012,Cochet-Escartin2014},
channel flow  \cite{Vedula2012,Marel2014},
confined flow \cite{Doxzen2013,Deforet2014},
collective migration \cite{Ashby2012,Vedula2013}. 
The dynamics of multicellular assemblies is regulated through 
mechanical forces that act upon cell adhesive structures.
These forces are exerted at the cell-substrate interface
\cite{Ladoux2012}, but also through cell-cell junctions \cite{Liu2010}.
The transmission of stresses within multicellular assemblies is thus
important to understand collective movements, cell rearrangements and 
tissue homeostasis. Even though kinematic information is readily available, 
mechanical properties that rely on internal stress are less well understood.
Indeed a number of important biological questions,
such as the determination of the molecular mechanisms that underlie 
the transmission of force within a tissue \cite{Bazellieres2015},
necessitate a measurement of internal stresses.

Several internal force measurement methods have been proposed and 
implemented (see \cite{ForcesinTissues2015} for a recent review):
at the molecular scale, F\"orster resonance energy transfer 
\cite{grashoff2010measuring,borghi2012cadherin}; 
at the cell scale, microrheology  \cite{Kole2004,tanase2007magnetic};
at the tissue scale, liquid drops \cite{campas2014quantifying}, 
birefringence \cite{nienhaus2009determination}, or
laser ablation \cite{Rauzi2008,bonnet2012mechanical}.
Although one would ideally like to read out from data the 
spatio-temporal dependence of the full stress field, the above methods 
yield either  a local, sub-tissue scale measurement
\cite{grashoff2010measuring,borghi2012cadherin,Kole2004,tanase2007magnetic,campas2014quantifying};  
or a subset of the components of the stress tensor 
\cite{nienhaus2009determination},
or a relative measurement, up to an undetermined multiplicative constant 
\cite{Rauzi2008,bonnet2012mechanical}. 
Monolayer stress microscopy (MSM), first introduced in  \cite{Tambe2011}, 
does not suffer from these drawbacks: it builds upon the measurement 
of traction force data to estimate the stress field of monolayers of
cohesive cells. 
Indeed, the force exerted by cells on a planar deformable substrate can be 
computed from the displacement field of the underlying layer
\cite{ambrosi2009traction,Style2014}, using either: 
\emph{(i)} traction force microscopy \cite{Dembo1999,Butler2002,Trepat2009}
where small beads are inserted within the (elastic) substrate,
their displacements are measured, and 
the traction forces are obtained by solving an inverse elastic problem;
or \emph{(ii)} arrays of micropillars 
\cite{Sniadecki2007,du2005force,saez2010traction}, 
where the traction forces are simply proportional to the in-plane 
displacements of the pillars.
However, once the traction forces are known, obtaining the internal stress 
from the force balance equations is an underdetermined problem since,
in the two dimensional case, three components of the 
symmetrical stress tensor must be obtained from two traction force components.
In MSM \cite{Tambe2011,Tambe2013},
these equations become well-posed thanks to an additional hypothesis 
on tissue rheology: the cell monolayer is assumed to be
a linear, isotropic elastic body. 
MSM has been validated independently on numerical data using 
particle dynamics simulations: in \cite{Zimmermann2014}, 
the reasonable accuracy of stress reconstruction from data that does not 
correspond to an elastic rheology has been attributed to the weakness of shear
stresses in both simulated and living tissues. Assuming again
that the tissue is an elastic body, and in addition that the displacement 
field is continuous at the cell-substrate interface,
internal stresses may also be computed directly
from substrate displacement data, circumventing the need to compute
traction forces \cite{Moussus2014}.

In the presence of cell divisions and extrusions that constantly
rearrange a tissue \cite{Ranft2010}, it is not clear that 
its rheology is that of a solid body. To our knowledge, 
the elastic rheology  hypothesis has not been directly validated, 
while alternative rheologies have been proposed in the 
literature \cite{Arciero2011,Lee2011,Koepf2013,Cochet-Escartin2014} 
and shown to model successfully specific aspects of the mechanical 
behaviour of cell monolayers.
Further, the rheology of multicellular assemblies may depend on the 
timescale \cite{Ranft2010}, as well as on the type of
cell considered \cite{Vedula2014}. These caveats call for a 
method to accurately estimate the internal stress field of a cell monolayer
irrespective of the underlying rheology.

A classical way to solve underdetermined inversion problems  
involves Bayesian inference \cite{kaipio2006statistical}, a technique 
originating in statistics \cite{Stuart2010}, and now 
widely used in physics \cite{vonToussaint2011} and 
biophysics \cite{kimura2015estimating}. Of note, Bayesian inversion 
has also been used to solve the inverse elastic problem of traction 
force microscopy \cite{dembo1996imaging,soine2015model}.
Recently, some among us proposed a Bayesian force inference method based on 
cell geometry, and applied it to segmented images of the Drosophila 
pupal wing and notum \cite{ishihara2012bayesian,Sugimura2013,Ishihara2013}. 
The tissue-scale stress arises from coarse-graining of cell-cell 
interactions. For tight epithelia where adherens junctions are a key player of 
force transmission between neighboring cells, it is reasonable to assume 
that the cell-scale contribution to stress is mostly related to 
local contact within the apical side of the epithelium, 
whereas basal contributions from, \emph{e.g.} lamellipodia, are negligible.
Accordingly, the dominant contributors to tissue-scale stress
were identified as cell pressures and cell-cell junction tensions, 
and force balance equations were written at each cell vertex, resulting 
in an  underdetermined system. This system was solved using
Bayesian inversion \cite{kaipio2006statistical}, where
the inferred tensions and pressures were the  most likely values (the modes) 
of a posterior distribution function.
In the case of the fruitfly pupal wing, it turned out that tissue
stress, obtained by coarse-graining, is oriented by 
external forces, and that its anisotropy promotes hexagonal
cell packing \cite{Sugimura2013}.
Similar systems of equations may become well-posed thanks to 
additional hypotheses  (equal cell pressures \cite{Ishihara2013,Chiou2012}), 
or when cell pressures are not required
\cite{Brodland2014}. However the stress is measured up to an 
arbitrary additive constant: its absolute value 
is out of reach since the input data  are cell vertex positions and 
cell junction angles. 

Below, we formulate Bayesian inversion stress microscopy (BISM), 
a method to estimate the internal stress 
field of a cell monolayer from traction force microscopy measurements.
Importantly, BISM yields an absolute measure of the stress and
dispenses with hypotheses on monolayer rheology.
We define BISM and introduce statistical measures of its accuracy. 
The method is first validated using numerical simulations that provide
traction force data. The inferred stress field, once computed,
is compared to the simulated stress data used as a reference.
Robustness is checked by implementing changes in the statistical model, 
as well as in the mechanical ingredients of the numerical simulations. 
BISM is further validated using experimental data 
in a quasi-one-dimensional geometry that allows for a direct calculation 
of the stress field by spatial integration of the traction force field. 
Finally, our results are 
compared with existing methods.

\section{Methods}
\label{sec:method}

\subsection{Mechanics}
\label{sec:method:mechanics}

\begin{figure}[!t]
\showfigures{
\includegraphics[scale=0.4]{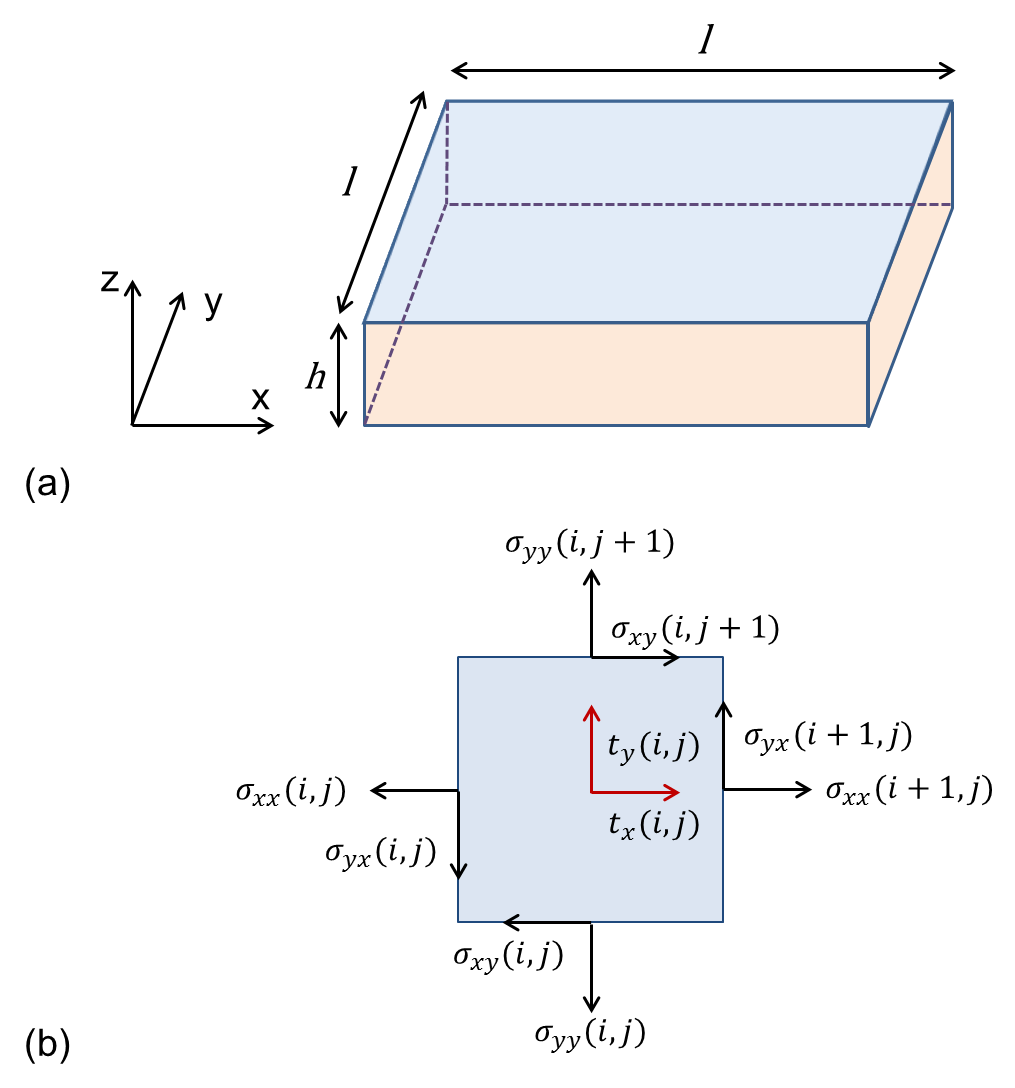}
}
\caption{
\textbf{Discrete monolayer mechanics.} 
(a) A tissue element of volume $\dx^2 \times h$.
(b) Pictorial representation of local force balance, 
Eqs.~(\ref{eq:discreteforce:x},\ref{eq:discreteforce:y}), see text for 
definitions. To see this figure in color, go online.}
\label{fig:force:balance}
\end{figure}

Within a continuum description, a flat, thin cell monolayer is
characterized at position $\vec{r}$ and time $t$ 
by a field of two-dimensional internal stresses $\sig(\vec{r},t)$ 
and by a field of external surfacic forces $\vec{t}(\vec{r},t)$
that the monolayer exerts on the  substrate. 
Since inertia is negligible,
the balance of linear momentum reads 
in vector form 
\begin{equation}
  \label{eq:forcebalance}
\mathrm{div} \, \sig= \vec{t},  
\end{equation}
and in cartesian coordinates $(x,y)$ 
(see \emph{Supplementary Text 1.2} for polar coordinates)
\begin{eqnarray}
\label{eq:force:x}
\DP{\sxx}{x}+\DP{\sxy}{y} &=& \Tx \\
\label{eq:force:y}
\DP{\syx}{x}+\DP{\syy}{y} &=& \Ty
\end{eqnarray}
Note that at this stage, due to the grid definition 
(see below and Fig.~\ref{fig:force:balance}b),
we do not enforce the symmetry of the stress tensor   
(equality of the shear stress components due to angular momentum 
conservation \cite{landau1975elasticity}). With a confined monolayer
in mind \cite{Doxzen2013,Deforet2014}, the boundary condition reads
\begin{equation}
  \label{eq:BC:general}
 \sigma_{ij} \, n_j=0  
\end{equation}
where $\vec{n}$ denotes the vector normal to the edge, and 
summation over repeated indices is implied.
In the plane, the units of stresses and (surfacic) forces are Pa.m
and Pa respectively.  Assuming that the monolayer height  is uniform and 
constant $h(\vec{r},t) = h_0$, the 3D stress reads 
$\sig_{3\mathrm{D}}= \sig/h_0$.
When spatial or temporal variations of the height cannot be neglected
\cite{Poujade2007,Trepat2009}, 
BISM can be implemented by replacing $\vec{t}(\vec{r},t)$ by 
$\vec{t}(\vec{r},t)/h(\vec{r},t)$ and by inferring the 3D stress
from $\mathrm{div} \, \sig_{3\mathrm{D}}= \vec{t}/h$,
provided that the height remains small compared to the system size,
as is generally the case for \emph{in vitro} cell monolayers
\cite{Tambe2011,zehnder2015thickness}.
A treatment of the full 3D case where the height is comparable
or larger than the system size is beyond the scope of this work.

Since experimental traction forces are measured with a finite spatial 
resolution $\dx$, assumed to be isotropic for simplicity,
we write a force balance equation in each of 
a large number of square surface elements of area $\dx^2$ 
(see Fig.~\ref{fig:force:balance}a).
We aim at inferring the stress tensor 
$\sig(i,j) = \bigl(\begin{smallmatrix}
\sigxx{i}{j} & \sigyx{i}{j} \\
   \sigxy{i}{j} & \sigyy{i}{j}
\end{smallmatrix} \bigr)$
in each element.
The traction force exerted by the tissue in element $(i,j)$ on 
the substrate is $\vT{i}{j}$, with components $\Txij{i}{j}, \Tyij{i}{j}$.
In the case of a rectangular grid with $C$ columns and $R$ rows, 
the discretized force balance equation for element $(i,j)$ reads
\begin{eqnarray}
  \label{eq:discreteforce:x}
\dx \, [\sigxx{i+1}{j} - \sigxx{i}{j} + \sigxy{i}{j+1} - \sigxy{i}{j}] 
 &=& \dx^2 \, \Txij{i}{j} \\
  \dx \, [\label{eq:discreteforce:y}
\sigyx{i+1}{j} - \sigyx{i}{j} + \sigyy{i}{j+1} - \sigyy{i}{j}]
 &=& \dx^2 \, \Tyij{i}{j}
\end{eqnarray}
to lowest order in $\dx$ (see Fig.~\ref{fig:force:balance}b).
We thus have $N = C \times R$ variables for $\Tx$ and $\Ty$, 
$(C+1) \times R$ variables for $\sxx$ and $\syx$ and $C \times (R+1)$ 
variables for $\syy$ and $\sxy$. 
Defining traction force and stress \emph{vectors} as
\begin{eqnarray*}
\vec{T}& = & [ \Txij{1}{1} \cdots \Txij{R}{C} \, \Tyij{1}{1} \cdots 
\Tyij{R}{C} ]^t\\
  \vec{\sig} &=&  [ \sigxx{1}{1} \cdots \sigxx{C+1}{R} \, 
\sigyy{1}{1} \cdots \sigyy{C}{R+1} \, \\
&&\sigxy{1}{1} \cdots \sigxy{C}{R+1} \, 
\sigyx{1}{1} \cdots \sigyx{C+1}{R}]^t
\end{eqnarray*}
where the superscript $^t$ denotes the transpose, we rewrite
Eqs.~(\ref{eq:discreteforce:x}-\ref{eq:discreteforce:y}) in matrix form
\begin{equation}
  \label{eq:Newton:disc}
 A \, \vec{\sig} = \vec{T} \,.
\end{equation}
The matrix $A$, of size $2N \times (4N+2(C+R))$, may be decomposed as
 \begin{equation}
 \label{eq:def:A}
 A=\bigl(\begin{smallmatrix}
 A_x & 0 & A_y & 0\\
  0& A_y & 0 & A_x
 \end{smallmatrix} \bigr)
 \end{equation}
where $A_x$ and $A_y$ correspond to the discretized matrix forms,
at second order in $\dx$, of the partial derivatives
with respect to $x$ and  $y$.

\subsection{Statistics}
\label{sec:method:statistics}

To solve the underdetermined linear system (\ref{eq:Newton:disc}), we
implement Bayesian inversion 
\cite{kaipio2006statistical}: all variables and parameters 
of the problem are probabilized. For simplicity, we use wherever 
possible Gaussian probability distribution functions,
denoted $\mathcal{N}(\vec{X} \mid \vec{m}, S)$ for a multivariate (vector)
Gaussian random variable $\vec{X}$ 
with mean $\vec{m}$ and covariance matrix $S$.

\paragraph{Likelihood}

The first ingredient of the statistical model 
is the likelihood function $L(\vec{T} \mid \vec{\sig})$,
which contains information provided by experimental measurements. 
For experimental data,
the force balance equations (\ref{eq:Newton:disc}) are  verified 
up to an additive noise due to measurement errors. Assuming this noise
to be Gaussian with zero mean and uniform covariance matrix $S=s^2 \, I$,
where the parameter $s^2$ denotes the noise variance and 
$I$ is the identity matrix, the likelihood is expressed as 
$L(\vec{T} \mid \vec{\sig})=\mathcal{N}(\vec{T} \mid A \vec{\sig}, s^2 \, I)$ or
\begin{equation}
\label{eq:likelihood}
L(\vec{T} \mid \vec{\sig}) = \left(\frac{1}{\sqrt{2\pi s^2}}\right)^{2N} \,
\mathrm{exp}\left[-\frac{\| \vec{T} - A \vec{\sig} \|^2}{2s^2} \right]
\end{equation}
where $\|\ldots\|$ is the (L$_2$) Euclidean norm.

\paragraph{Prior}

Second, the prior probability distribution function $\pi(\vec{\sig})$
embeds additional information concerning the stress field:
\begin{itemize}
\item[\emph{(i)}]{we assume that the stress obeys a Gaussian distribution function  with  zero mean
$\vec{\sig}_0= \vec{0}$ and covariance matrix $s_0^2 \, I$;}
\item[\emph{(ii)}]{we enforce the equality of the two off-diagonal 
components of the stress tensor: in compact vector form 
$\vec{\sigma}_{xy} =\vec{\sigma}_{yx}$, \emph{i.e.} 
$\sigxy{i}{j}+\sigxy{i}{j+1} = \sigyx{i}{j}+\sigyx{i+1}{j} \;\; 
\forall (i,j)$ (see Fig.~\ref{fig:force:balance}b and \emph{ST 1.3});}
\item[\emph{(iii)}]{
we enforce the boundary conditions
(\ref{eq:BC:general}), namely two conditions at each boundary element $(i,j)$, 
written in compact vector form $\vec{\sig}_{\mathrm{BC}} = \vec{0}$).}
\end{itemize}
Up to a normalizing factor, the prior reads
\begin{equation}
\label{eq:prior}
\pi(\vec{\sig}) \propto 
\mathrm{exp} \left[ -\frac{  \|  \vec{\sig}   \|^2 + 
\axy^2 \| \vec{\sigma}_{xy} -\vec{\sigma}_{yx} \|^2 + 
\aBC^2 \| \vec{\sig}_{BC} \|^2 }{2s_0^2} \right]
\end{equation} 
or 
\begin{equation}
\label{eq:prior:reduced}
\pi(\vec{\sig}) =
\left( 
\frac{1}{ \sqrt{2\pi  |B| s_0^2}} \right)^{4N + 2(R+C)} \,
\mathrm{exp} \left[ -\frac{\vec{\sig}^t B^{-1} \vec{\sig}}{2 s_0^2}   \right]
\end{equation} 
The second ingredient of the statistical model is a Gaussian prior
$\pi(\vec{\sig})=\mathcal{N}(\vec{\sig} \mid \vec{0}, S_0 = s_0^2 \, B)$
where $B$ is a reduced covariance matrix of determinant $|B|$.
Note that a Gaussian prior suppresses stress values larger than a 
few times $s_0$ (see \cite{schwarz2002regularization} for a similar
approach in the context of traction force microscopy).
In practice, we set the hyperparameters $\axy$ and $\aBC$ to the 
values $\axy=\aBC=10^3$, large enough for conditions \emph{(ii)} and 
\emph{(iii)} to be enforced
(see \emph{ST 3.2} for a discussion of these values). 
If required by a given experimental set-up, the boundary 
conditions should be modified appropriately in the definition of the prior.

\paragraph{Resolution}

According to Bayes' theorem, 
the posterior (conditional) probability distribution function 
$\Pi(\vec{\sig} \mid \vec{T})$ 
of the stress given the traction force data is proportional to the product of
the likelihood by the prior 
\begin{equation}
  \label{eq:Bayes}
 \Pi(\vec{\sig} \mid \vec{T}) \propto L(\vec{T} \mid \vec{\sig}) \times
\pi(\vec{\sig}) 
\end{equation}
Since both are Gaussian, the posterior is also Gaussian 
$\Pi(\vec{\sig} \mid \vec{T}) = \mathcal{N}(\vec{\sig} \mid 
\vec{\sig}_{\Pi}, S_{\Pi})$, with a covariance matrix $S_{\Pi}$
and a mean $\vec{\sig}_{\Pi}$ given by \cite{kaipio2006statistical}
\begin{eqnarray}
\label{eq:post:Gaussian:cov}
S_{\Pi} & = &  \left( S_0^{-1} + A^t \, S^{-1} \, A \right)^{-1}\\
\label{eq:post:Gaussian:mean}
\vec{\sig}_{\Pi} & = & 
S_{\Pi} \,  A^t \, S^{-1} \, \vec{T} 
\end{eqnarray}

We use maximum a posteriori (MAP) estimation
\cite{kaipio2006statistical} and define the inferred stress $\vec{\hat{\sig}}$
as the mode (maximal value)
of the posterior $\vec{\hat{\sig}} = \vec{\sig}_{\Pi}$. 
Qualitatively, the underdeterminacy has been lifted: $4N+2(R+C)$ 
unknown stress values are determined from $2N$ traction force values, 
$4N+2(R+C)$ conditions from the Gaussian distribution of the stress tensor, 
$N$ equalities of the two shear components and $4(R+C)$ boundary conditions. 

In this Gaussian model, MAP estimation is identical to minimization 
of a Tikhonov potential \cite{kaipio2006statistical}.
The dimensionless regularization parameter
\begin{equation}
  \label{eq:def:Lambda}
\Lambda = \frac{l^2 \,s^2}{ s^2_{0}}  
\end{equation}
quantifies the relative weight given to the prior, compared to
the likelihood, when performing Bayesian inversion. Factoring out $(sl)^2$, 
Eqs.~(\ref{eq:post:Gaussian:cov}-\ref{eq:post:Gaussian:mean}) read
 \begin{eqnarray}
 \label{eq:post:Gaussian:cov:Lambda}
 S_{\Pi} & = &  (sl)^2\left( \Lambda\,B^{-1} +l^2\, A^t \, A \right)^{-1}\\
 \label{eq:post:Gaussian:mean:Lambda}
 \vec{\sig}_{\Pi} & = & 
 \left( \Lambda\,B^{-1} +l^2\, A^t \, A \right)^{-1} l^2\, \,  A^t \, \vec{T} 
 \end{eqnarray}
Since the product $Al$ is dimensionless and independent of $l$, 
the posterior covariance (\ref{eq:post:Gaussian:cov:Lambda}) is a 
function of $\Lambda$ and $sl$, while the posterior mode 
(\ref{eq:post:Gaussian:mean:Lambda}) depends upon $\Lambda$ and $l\vec{T}$.

\begin{figure}[!t]
\centering 
\showfigures{
\includegraphics[scale=0.4]{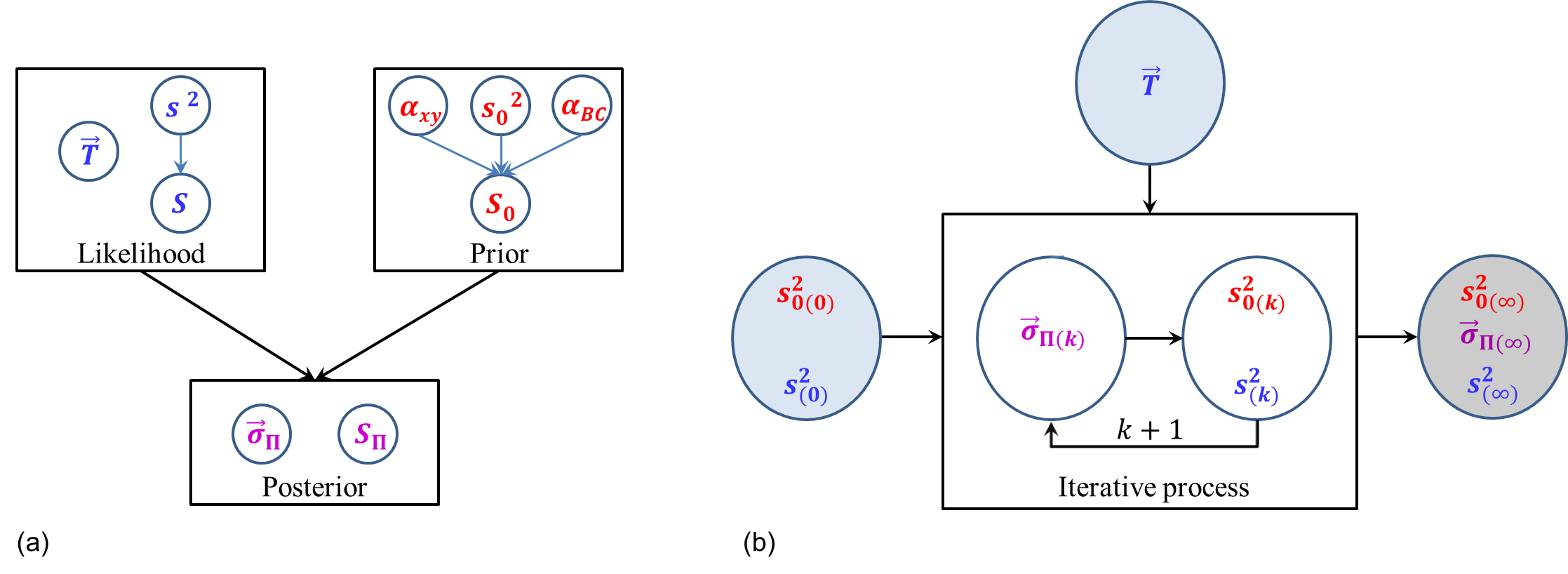}
}
\caption{\label{fig:hierarchical}
\textbf{Schematics of BISM.}
To see this figure in color, go online.
(a) Both the likelihood (with parameter $s^2$), and the prior
(with hyperparameters $s_0^2$, $\axy$ and $\aBC$) contribute
to the definition of the posterior, a Gaussian
distribution function of mean $\vec{\sig}_{\Pi}$ and covariance
matrix $S_\Pi$. Given numerical values of $s^2$, $s_0^2$, $\axy$ and $\aBC$, 
the MAP estimator of the stress $\vec{\hat{\sig}}$ is the mode of the posterior
$\vec{\sig}_{\Pi}$, Eq.~(\ref{eq:post:Gaussian:mean}).
(b) In BISM, the values of $s^2$ and $s_0^2$ are not given \emph{a priori},
but determined self-consistently within a hierarchical Bayesian construction.
MAP estimation is performed iteratively, by successively optimizing 
the posterior for the mode $\vec{\sig}_{(k)}$, 
Eqs.~(\ref{eq:post:Gaussian:cov}-\ref{eq:post:Gaussian:mean}),
and for the variances ${s^2}_{(k)}$, ${s_0^2}_{(k)}$, 
Eqs.~(\ref{eq:hyperparameters:s}-\ref{eq:hyperparameters:s0}),
until convergence to a fixed point is reached. The estimator of the stress 
$\vec{\hat{\sig}}$ is defined as the asymptotic value 
$\vec{\sig}_{\Pi \,\infty}$, computed using 
$s^2_{(\infty)} = \lim_{k \to \infty} {s^2}_{(k)}$ and 
$s^2_{0(\infty)} = \lim_{k \to \infty} s_{0 \, (k)}^2$ in
Eqs.~(\ref{eq:post:Gaussian:cov}-\ref{eq:post:Gaussian:mean}),
with a regularization parameter
$\Lambda_{(\infty)} = l^2 \,s^2_{(\infty)}/ s^2_{0\,(\infty)}$.
A flowchart of the algorithm is given in Fig.~S6 
in the Supporting Material.
}
\end{figure} 

\paragraph{Hyperprior}

For generality sake, we probabilize the parameter $s^2$ and 
the hyperparameter $s_0^2$, yet undetermined
in Eqs.~(\ref{eq:post:Gaussian:cov}-\ref{eq:post:Gaussian:mean})
(recall that  $S = s^2 \, I$ and $S_0 = s_0^2 \, B$).
Within the framework of hierarchical Bayesian descriptions, the model 
is closed by the \emph{hyperprior} probability distribution
functions $H(s^2)$ and $H(s_0^2)$ \cite{Carlin2008,gelman2014bayesian}.
Up to a normalizing factor, the posterior now reads
(see Fig.~\ref{fig:hierarchical}a)
\begin{equation}
  \label{eq:Bayes:hierarchical}
 \Pi(\vec{\sig} \mid \vec{T}) \propto L(\vec{T} \mid \vec{\sig}, s^2) 
\times \pi(\vec{\sig} \mid  s_0^2)  \times H(s^2) \times H(s_0^2)
\end{equation}
For simplicity, we use Jeffreys' non-informative hyperprior: 
$H(s^2) \propto  1/s^2$,  $H(s_0^2) \propto 1/s_0^2$ 
\cite{Carlin2008,gelman2014bayesian}.

Simultaneous a posteriori optimization with respect to $\vec{\sigma}$,
$s^2$ and ${s_0^2}$ being intractable, we solve the problem iteratively, 
starting from initial values  $s^2_{(0)}$ and ${s_0^2}_{(0)}$ 
(see Fig.~\ref{fig:hierarchical}b).
At step $k \ge 1$, we first calculate the mode $\vec{{\sig}}_{\Pi \,(k)}$ from 
previous values ${s^2}_{(k-1)}$, ${s_0^2}_{(k-1)}$,
Eqs.~(\ref{eq:post:Gaussian:cov}-\ref{eq:post:Gaussian:mean}). 
Maximizing the posterior with respect to each hyperparameter yields 
the updated hyperparameter values
\begin{eqnarray}
\label{eq:hyperparameters:s}
{s^2}_{(k)} &=&\frac{1}{2N+2} \; \| \vec{T} - A \vec{{\sig}}_{\Pi \,(k)} \|^2
\\
\label{eq:hyperparameters:s0}
{s_0^2}_{(k)}&=& \frac{1}{4N+2(R+C)+2} \;
\vec{{\sig}}_{\Pi \,(k)}^t \, B^{-1} \, \vec{{\sig}}_{\Pi \,(k)}
\end{eqnarray}
Once convergence is reached, ${s^2}_{(k)} \to s^2_{(\infty)}$, 
${s_0^2}_{(k)} \to s^2_{0(\infty)}$, the stress estimate  is defined 
as $\vec{\hat{\sig}} = \vec{\sig}_{\Pi \,(\infty)}$, computed from
Eqs.~(\ref{eq:post:Gaussian:cov}-\ref{eq:post:Gaussian:mean})
with the optimal values $s^2_{(\infty)}$ and $s^2_{0(\infty)}$.
An estimate $\vec{\delta \hat{\sig}}$ of the error on $\vec{\hat{\sig}}$
is calculated as the  square root of the diagonal values  of the 
covariance matrix $S_{\Pi \, (\infty)}$. Since the marginal distribution of 
traction forces is Gaussian, with a covariance matrix 
$S_T = S+A \, S_0 \, A^t$ \cite{kaipio2006statistical}, we also calculate an 
estimate $\vec{\delta \hat{T}}$ of the error on the traction force as the 
square root of the diagonal values of $S_T$.

\subsection{Measures of accuracy}
\label{sec:method:quant}

The numerical resolution of a set of hydrodynamical equations 
yields a numerical data set $\{t^{\mathrm{num}}\}$ of traction forces, from which
we compute a set $\{\sigma^{\mathrm{inf}}\}$
of inferred stresses. Since the numerical data set 
$\{ \sigma^{\mathrm{num}}\}$ of stresses  is also available, measures
of accuracy involving numerical simulations 
typically compare $\{\sigma^{\mathrm{inf}}\}$ with $\{ \sigma^{\mathrm{num}}\}$. 

A classical ``goodness-of-fit'' measure is the coefficient of determination,
defined for the $\sig_{xx}$ component of the stress as
\begin{equation}
  \label{eq:def:R2}
  R_{xx}^2 = 1-\frac{\sum (\sig_{xx}^{\mathrm{num}}-\sig_{xx}^{\mathrm{inf}})^2}
{\sum (\sig_{xx}^{\mathrm{num}}- \langle \sig_{xx}^{\mathrm{num}} \rangle)^2}
\end{equation}
where the sums and the averages $\langle \ldots \rangle$ 
are performed over space. Similar definitions apply to
other components, and allow to define an aggregate coefficient of 
determination $R^2_{\sig}$ averaged over all stress components.
Accurate estimates correspond to numerical values of $R_{\sig}^2$ close to $1$.
The discretized force balance equations 
(\ref{eq:discreteforce:x}-\ref{eq:discreteforce:y}), used as a definition
of inferred traction forces, yield a set $\{t^{\mathrm{inf}}\}$ of inferred 
traction forces computed from the set of inferred stresses
$\{\sig^{\mathrm{inf}}\}$. Comparing $\{t^{\mathrm{num}}\}$ with
$\{t^{\mathrm{inf}}\}$  allows to define similarly a $R^2_T$ diagnostic 
for numerical data.

When analyzing an experimental data set $\{t^{\mathrm{exp}}\}$ of 
traction forces, $R^2_{\sig}$ cannot be computed in the absence of a 
reference set of stresses. As above, comparing $\{t^{\mathrm{exp}}\}$ 
with $\{t^{\mathrm{inf}}\}$ allows
to define a measure of accuracy  for experimental data,
the coefficient of determination $R^2_T$.  
An alternative measure of predictive accuracy is the 
$\chi^2_{T}$ diagnostic, defined as the average value
of the square of reduced residuals
\begin{equation}
  \label{eq:def:chi:texp}
  \chi^2_T = \frac{1}{2N} 
\, \sum \frac{(t^{\mathrm{exp}}-t^{\mathrm{inf}})^2}{\delta \hat{T}^2}
\end{equation}
where the sum is performed over space and over traction force components.
This measure of accuracy is, up to a normalizing factor, similar
to the ``omnibus goodness-of-fit'' measure advocated in
\cite{Carlin2008,gelman2014bayesian}. 
The estimated standard deviation $\delta \hat{T}$
may be replaced in Eq.~\eqref{eq:def:chi:texp}
by the measurement error $\delta t$. 
Numerical values of $\chi^2_T$ close to $0$ are indicators of high accuracy.

A last test of accuracy is provided by the calculation of average 
experimental stress values from traction force data 
\cite{landau1975elasticity}, which may be compared with the average 
inferred stresses for each component (see Sec.~\ref{sec:results:exp} and 
\emph{ST 1.4}).

\subsection{Experimental methods}
\label{sec:method:exp}

We used MDCK (Madin-Darby canine kidney) cells as an epithelial cell model.

\paragraph{Cell Culture}

MDCK wild-type cells were cultured in media containing DMEM (Life
Technologies), $10 \, \%$ FBS (Life Technologies) and $1 \, \%$ 
antibiotics (penicillin and streptomycin).

\paragraph{Micro-contact printing and substrate 
preparation  for  Traction  Force Microscopy}

We measured the traction forces exerted  by  cells  on  their  substrate  by
using soft silicone gel as previously described \cite{Vedul2014}. 
Fluorescent beads  were deposited onto the gel to measure the displacement 
field.  Briefly,  a  thin layer of the gel was spread on a glass bottom 
dish and then  cured  at  $80^o$C for $2$ hours. Cured gel was silanized 
using $5\%$  solution  of  (3-aminopropyl) triethoxysilane  (APTES,  Sigma)  
in  pure  ethanol.  This  gel  was   later incubated  for  $5$  minutes  
with  $100$ nm  carboxylated  fluorescent   beads (Invitrogen) suspended 
in deionized (DI) water. Subsequently, the  substrate was dried and 
micro-contact printed with fibronectin \cite{Vedula2013,Fink2007}  
using  a  thin water soluble Polyvinyl Alcohol (PVA) membrane which 
allows the transfer  of fibronectin on soft gel. The PVA membrane was 
later dissolved and  the  non-contact printed areas were blocked using 
$0.2 \, \%$  pluronics  (Sigma)  solution. The substrate was then washed 
and was seeded with cells. Cells were  allowed to grow until the 
micro-contact printed area was fully covered.

The images were acquired using phase contrast and  fluorescent  channels  to
record cell positions and bead displacements, respectively.

For analysis, the imaging drifts were corrected in \texttt{ImageJ}  (NIH)  
using  the Image Stabilizer plugin \cite{Li2008}. To analyze the 
displacement field of beads,  we used an open source iterative PIV 
(particle  image  velocimetry)  plugin  in \texttt{ImageJ} \cite{Martiel2015}. 
To reconstruct  the  traction  force  field  from  the  obtained
displacement field, an open  source  Fourier  transform  traction  
cytometry (FTTC) plugin was used in \texttt{ImageJ} \cite{Martiel2015}. 
The resulting  traction  force  values were taken for the  
validation  of  the  BISM  inferred  stress fields.
To estimate the experimental error $\delta t_{\mathrm{exp}}$ 
made on traction force measurements, we calculated the mean value
of traction forces measured on square regions of the substrate devoid of cells
(surface area $50 \times 50 \, \mu\mathrm{m}^2$).

\paragraph{Monolayer height measurement}
The confluent cell monolayer was fixed for immunofluorescence microscopy. 
Actin present inside the cells was fluorescently labeled using 
Alexa Fluor 488-conjugated phalloidin (Invitrogen) 
at 1:1000 dilution  in PBS. To measure tissue height, overall cell
shape was then visualized with the help of cortical actin. Imaging was done
using a Zeiss LSM 780 confocal  microscope with a step size 
of $0.4 \, \mathrm{\mu m}$ to capture the entire height of the
tissue. In the confined ring shape geometry, 
height was measured at different locations to obtain the 
mean value and standard deviation of the monolayer height
$h = 5.3 \pm 1.2 \, \mu$m.

\section{Results}
\label{sec:results}

\subsection{Validation: numerical data} 
\label{sec:results:numerical}

\begin{figure}[!t]
\centering 
\showfigures{
\includegraphics[scale=0.35]{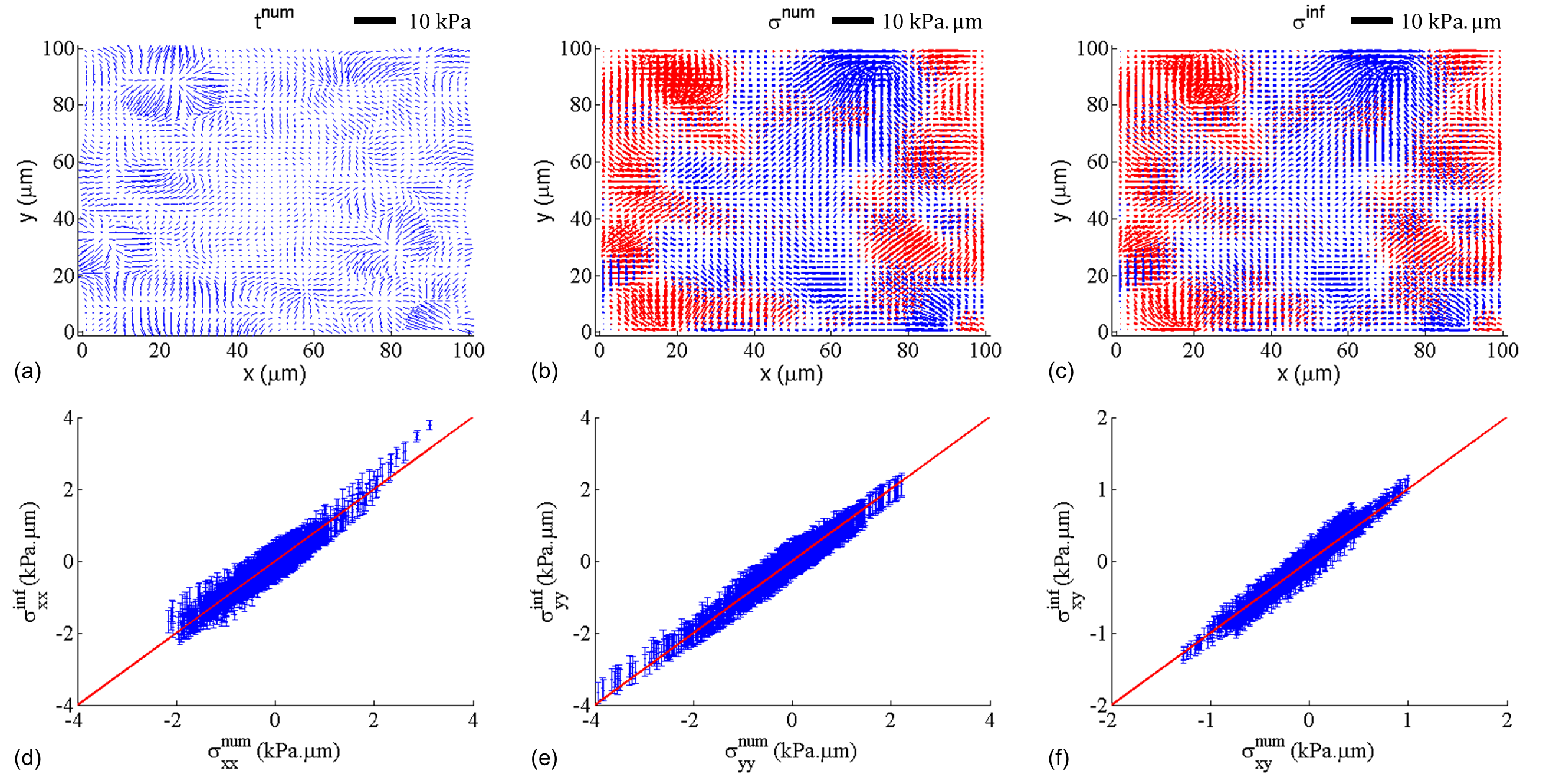}
}
\caption{\label{fig:validation:numerical}
\textbf{Validation: numerical data.} 
(a) Simulated traction force field $t^{\mathrm{num}}$, represented at each point
by an arrow. Scale bar: $10$ kPa.
(b) Simulated stress field $\sigma^{\mathrm{num}}$.
(c) Inferred stress field $\sigma^{\mathrm{inf}}$,
plotted on a $50 \times 50$ grid. 
At each point, the stress tensor is represented by two line segments
oriented along the stress eigenvectors, of lengths proportional
to the eigenvalues (blue: tensile stress, red: compressive stress, 
scale bar: $10$ kPa $\mu$m).
Note the high degree of similarity between images (b) and (c).
(d-f) Plots of the inferred stress \emph{vs.} the simulated stress 
for each component, in kPa.$\mu$m. Error bars correspond to 
$\delta \hat{\sigma}$ and the red line is the bisector $y = x$.
To see this figure in color, go online.
}
\end{figure}

A first example of the application of BISM to a numerical data set
is given by using the traction force field of a compressible viscous fluid 
driven by active force dipoles, interacting with its substrate through 
an effective fluid friction force, and confined in a square,
with the boundary conditions (\ref{eq:BC:general}). 
We solve this problem on a $100 \times 100 \, \mu\mathrm{m}^2$ square 
over a regular cartesian grid with $C=R=50$, $N=C\times R=2500$ and
 \mbox{$l=2 \, \mu\mathrm{m}$} (see \emph{ST 1.1}).
We use material parameter values typical of cell monolayers:
friction coefficient 
$\xi_v=10^0 \,\mathrm{kPa\, \mu m}^{-1}\mathrm{s}$ \cite{Cochet-Escartin2014}, 
shear viscosity $\eta=10^{3} \, \mathrm{kPa \, \mu m\, s}$ \cite{Harris2012}, 
and compression viscosity $\eta'= \eta$. 
To account for the measurement error, we add to the traction force field
a white noise of relative amplitude $5 \%$  
(variance $s_{\mathrm{exp}}^2=1.2\;10^{-3} \, \mathrm{kPa^2}$),
and obtain the numerical data set $\{t^{\mathrm{num}}\}$ of traction forces (Fig.~\ref{fig:validation:numerical}a),
referred to below as \emph{Viscous}.
We checked that the total sum of the traction forces is close to zero,
as expected for a closed system with negligible inertia.

Bayesian inversion is performed with a custom-made script
written in \texttt{Matlab} (The MathWorks, Inc.).
With ${s^2}_{(0)}=10^{-1} \, \mathrm{kPa^2}$ and 
${s_0^2}_{(0)}=10^2 \, \mathrm{kPa^2 \,\mu m^2}$ as initial values,  
the resolution method converges in a few steps towards the asymptotic 
hyperparameter values ${s^2}_{(\infty)}=4.7\, 10^{-7} \, \mathrm{kPa^2}$ and 
${s_0^2}_{(\infty)}=3.4\, 10^{-1} \, \mathrm{kPa^2 \,\mu m^2}$,
or $\Lambda_{(\infty)} = 5.5 \, 10^{-6}$ 
(see \emph{ST 3.1} and Fig.~S4a in the Supporting Material) 
The data sets  $\{\sigma^{\mathrm{num}}\}$ and $\{\sigma^{\mathrm{inf}}\}$
of simulated and inferred stresses
are shown in Fig.~\ref{fig:validation:numerical}b-c: we find
that their spatial structures are quite similar. This observation is confirmed 
quantitatively by plotting component by component the inferred stress 
\emph{vs.} the simulated stress (Figs.~\ref{fig:validation:numerical}d-f), 
and by the numerical values of the coefficients of determination
$R^2_{xx}=0.94$, $R^2_{yy}=0.97$, $R^2_{xy}=0.95$, 
yielding an aggregate measure of accuracy $R^2_{\sig} = 0.96$, close to $1$.  
Using the inferred data set $\{t^{\mathrm{inf}}\}$ of traction forces,
we also obtain $1 - R^2_T = 2\,10^{-5}$ and $\chi^2_T=7\;10^{-7}$, indicating 
that all the information contained in the traction force data is used. 
An order of magnitude of the error bar is given by the standard deviations,
calculated using (\ref{eq:post:Gaussian:cov:Lambda}): 
$\delta\hat{\sig} \approx 10^2\,ls_{(\infty)} \approx 10^{-1} \, 
\mathrm{kPa\, \mu m}$, corresponding to $\approx 10 \,\%$ of the 
maximal stresses.

All qualitative and quantitative indicators show that the stress field
has been inferred accurately.

\subsection{Robustness to variations of the statistical model}

We test the robustness of BISM by varying
\emph{one by one} each feature of the statistical model,
first focusing on alternative definitions of the prior, 
arguably our most prominent assumption, second modifying
the likelihood, the hyperprior and the resolution method.
For conciseness, precise definitions and implementations are given 
in \emph{ST 2.1-2.3}. Table~S1 in the Supporting Material
lists the values of $R^2_{\sig}$ thus obtained,  given the same 
numerical data sets as for BISM.

Setting $\axy$ to $0$ in the definition of the prior
has a significant influence on the accuracy of inference
($R^2_{\sig} = 0.75$): the symmetry of the stress tensor needs to be 
enforced in the prior for accurate estimation. Unsurprisingly, 
this impacts less the diagonal  
($R^2_{xx}=R^2_{yy}=0.81$) than the shear components ($R^2_{xy}=R^2_{yx}=0.61$).
In a similar way, knowledge of the correct boundary conditions should
be included in the prior whenever possible: setting $\aBC$ to $0$
has a large negative impact $(R^2_{\sig} = 0.53$).
We shall further comment below on the influence of boundary conditions.

Importantly, the accuracy of inference
remains excellent when the prior, the likelihood, or the hyperprior 
distributions are not Gaussian.
This shows that the accuracy of BISM does not depend sensitively 
on a Gaussian assumption.
Of note, we do not assume that traction force \emph{data} obeys
a Gaussian distribution. The data set ($\{t^{\mathrm{num}}\}$ or 
$\{t^{\mathrm{exp}}\}$) is used \emph{as is}  -- indeed experimental 
traction force distributions are known to exhibit exponential 
tails \cite{Trepat2009,gov2009traction}.
In all cases, the regularization parameter is small
$\Lambda_{(\infty)} \ll 1$ (see Table~S3 in the Supporting Material):
the distribution of inferred stresses $\{\sigma^{\mathrm{inf}}\}$ depends 
mostly on the empirical distribution of traction force data.
Thus, even if the multivariate posterior distribution is Gaussian, 
the univariate, empirical distribution of the inferred stress 
(the mode $\vec{\sig}_{\Pi\,(\infty)}$) is not necessarily Gaussian
(see Fig.~S7 in the Supporting Material for numerical data).
Similarly, even if the stress prior distribution function has
a zero mean, the mean inferred stress is not necessarily equal to zero
(see Fig.~\ref{fig:validation:exp}i for an example).
The small values of $\Lambda_{(\infty)}$ are consistent with robustness
with respect to variations of the prior. 

We conclude that BISM is robust to variations of the statistical model.

\begin{figure}[!t]
\centering 
\includegraphics[scale=0.3]{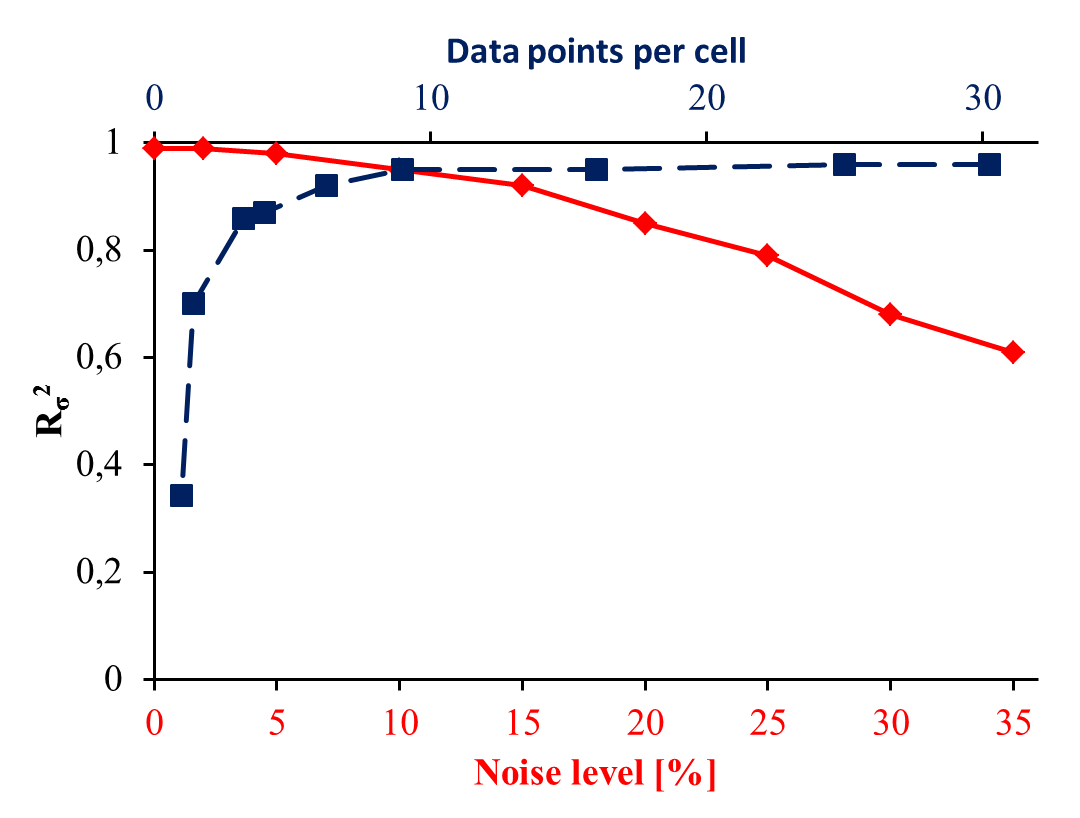}
\caption{\label{fig:robustness}
\textbf{Robustness.}
BISM remains highly accurate ($R^2_{\sig} > 0.8$) for noise levels 
and spatial resolutions typical of traction force measurements.
Red solid line: coefficient of determination $R^2_{\sig}$ 
\emph{vs.} relative level of added noise. 
$R^2_{\sig}$ is averaged over $3$ realizations of the noise.
Blue dashed line: coefficient of determination $R^2_{\sig}$ 
\textrm{vs.} number of traction force data points per cell, 
for a typical cell area of $100 \, \mathrm{\mu m^2}$. 
To see this figure in color, go online.
}
\end{figure}

\subsection{Robustness to variations of the numerical simulation}
\label{sec:results:robust:sim}

We next apply BISM to a broad spectrum of numerical data, and vary 
successively the values of material parameters, the rheology, 
the boundary conditions, the system shape, the spatial resolution, 
the nature and the amplitude of the measurement noise
(see \emph{ST 2.2}).

Given the values of $R^2_{\sig}$ compiled in Table~S2 in the
Supporting Material, we verify that BISM remains accurate with a different 
(elastic) rheology (also set \emph{Elastic 1} in Table~\ref{tab:MSM}), or 
in a different (circular) geometry.  In the viscous case, we observe that 
the accuracy decreases as a function of the bulk viscosity $\eta'$.
For larger $\eta'$, we observe larger values of stress components, 
leading to  asymmetrical distributions: this is at variance with our 
assumption of an even prior distribution, and may explain the lower value of
$R^2_{\sig}$ obtained when $\eta' = 10^{1} \, \eta$.

Accuracy decreases when the prior does not include 
our knowledge of the boundary conditions.
However the influence of erroneous values of the stress
at the system's boundaries rapidly decreases far from the edge.
When the coefficient $\aBC$ is set to $0$, removing the outtermost rows and 
columns  increases $R^2_{\sig}$ from $0.53$ to $0.60$. 
We estimate the corresponding 
``penetration length'' of boundary values to about $10 \%$ of the 
system size, consistent with \cite{Tambe2013}. 
We conclude that, whenever available, the correct boundary 
conditions should be taken into account in the prior.

Importantly (see Fig.~\ref{fig:robustness}),
accuracy remains acceptable ($R^2_{\sig} > 0.8$) for 
a spatial resolution larger than a few data points per cell, as well as for 
measurement noise levels up to $20 \, \%$  of the traction force amplitude,
consistent with measurement errors typical of 
force traction microscopy \cite{Sniadecki2007}.

These results highlight that BISM is robust to variations of the numerical
simulations that yield the traction force data.

\subsection{Validation: experimental data}
\label{sec:results:exp}

\begin{figure}[!t]
\begin{adjustwidth}{-1.0in}{0in}
\centering
\showfigures{
\includegraphics[scale=0.39]{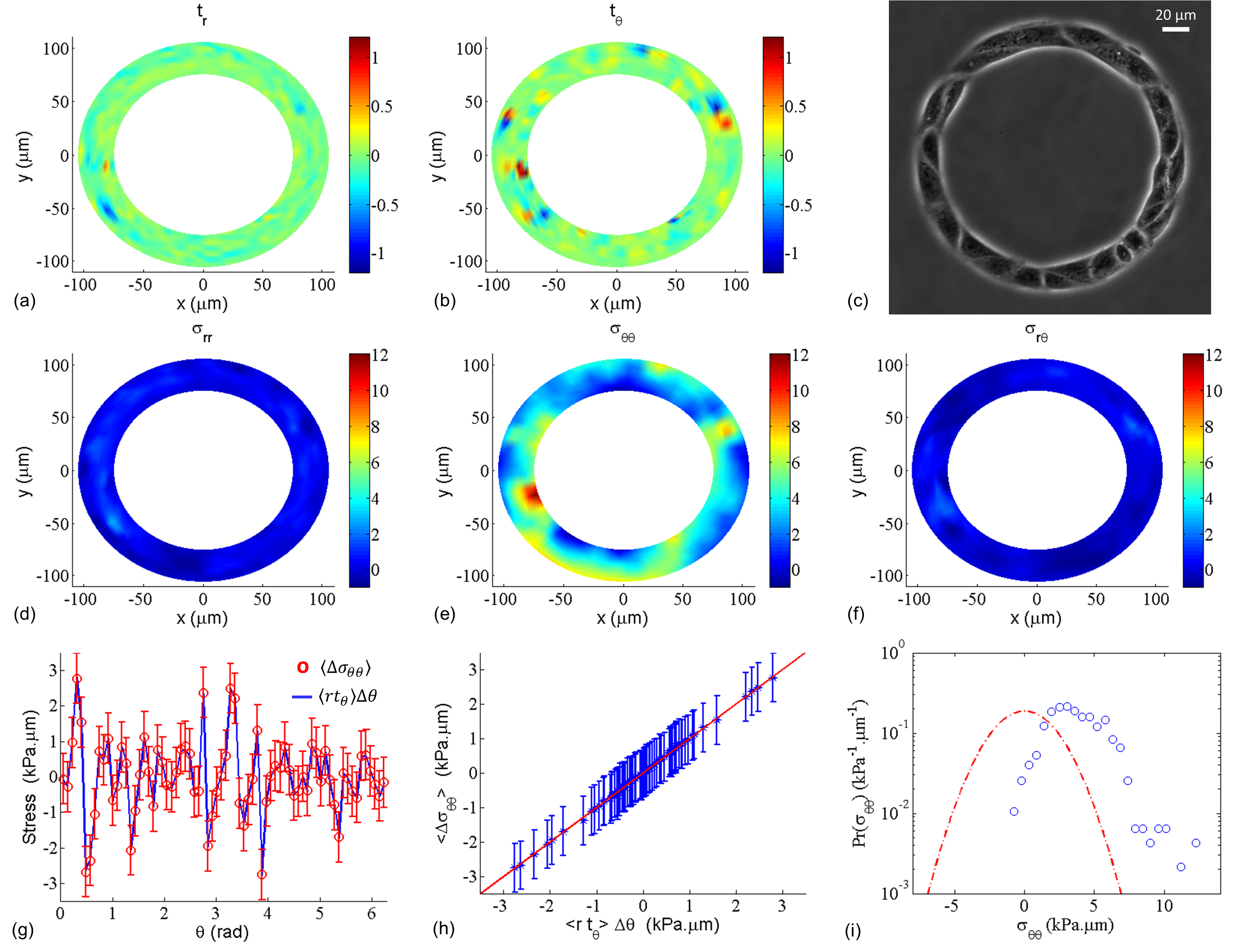}
}
\caption{
\label{fig:validation:exp}
\textbf{Validation: experimental data.}
(a-b) Heat maps of the components $t_r$ and $t_{\theta}$ of the traction forces 
$t^{\mathrm{exp}}$ in kPa, on a $12 \times 72$ polar grid. 
(c) Phase constrast image of the MDCK cell monoloayer. Scale bar: $20 \, \mu$m.
(d-f) Heat maps of the components $\sig_{rr}$, $\sig_{\theta\theta}$ and 
$\sig_{r\theta}$ of the inferred stress field $\sig^{\mathrm{inf}}$ in 
kPa$\,\mu$m. 
(g) Angular profiles of the radially-averaged inferred stress 
($\langle \Delta \sig_{\theta\theta} \rangle_r$, red circles) 
and of the 1D stress ($\langle r \, \Tt \rangle_r \, \Delta \theta$,
blue line), with an angular resolution $\Delta \theta = \pi/36$ rad. 
Error bars of the inferred stress are 
the radial average of $\delta \hat{\sigma}_{\theta \theta}$.
(h) Radially-averaged inferred stress \emph{vs.} the 1D stress. 
The coefficent of determination of this plot is $R^2_{\mathrm{ring}}=0.99$. 
(i) Empirical distribution function of the inferred 
component $\stt$ (blue circles). The red dashed line corresponds 
to the zero-mean, Gaussian prior distribution function
with standard deviation $s_0=2.13 \, \mathrm{kPa.\mu m}$. 
To see this figure in color, go online.
}
\end{adjustwidth}
\end{figure}

While inverting  the force balance equations (\ref{eq:forcebalance}) 
requires specific techniques in 2D, the same problem reduces in 1D  
to straightforward integration along the spatial coordinate \cite{Trepat2009}.
For this reason, we fabricated
a micro-patterned ring whose measured mean radius 
$r_{\mathrm{mean}}=90 \,\mu \mathrm{m}$ is larger than its width 
$w=33 \, \mu \mathrm{m}$, measured the substrate displacement 
field, and deduced the traction forces exerted
by a  monolayer of MDCK cells confined within the ring
(Figs.~\ref{fig:validation:exp}a-c, see \emph{Experimental Methods} for 
further details). We find an average traction force amplitude 
$t_{\mathrm{exp}} \approx 200$ Pa for a measurement error
of the order of $\delta t_{\mathrm{exp}} \approx 40$ Pa, and deduce 
a relative error $\delta t_{\mathrm{exp}}/t_{\mathrm{exp}} $
of the order of $20 \,\%$, consistent with the range of noise amplitudes 
where BISM was deemed applicable, see Fig.~\ref{fig:robustness}. 
The height of the monolayer is typically $5.3 \pm 1.2 \, \mu$m
(see Sec.~\ref{sec:method:exp}),
much smaller than the spatial extension $2 \pi r_{\mathrm{mean}}$,
and varies smoothly, see Fig.~S8 in the Supporting Material.
In Figs.~\ref{fig:validation:exp}d-f, we plot the three stress components 
as inferred by BISM, with a regularization parameter 
$\Lambda_{(\infty)} = 6.7 \, 10^{-6}$
and a traction force-based measure of accuracy 
$\chi_T^2=4.7 \, 10^{-6}$, as defined by Eq.~(\ref{eq:def:chi:texp}).
Note that the inferred stresses are mostly positive, even though 
the prior distribution is a zero-mean Gaussian, 
see Fig.~\ref{fig:validation:exp}i.

Since shear stresses are small compared to angular normal stresses
$|\srt|, |\str| \ll |\stt|$, the orthoradial component of the 
force balance equation (see \emph{ST 1.2})  simplifies to
\begin{equation}
  \label{eq:force:t:red}
  \DP{\stt}{\theta} = r \, \Tt
\end{equation}
Taking into account the experimental angular resolution 
$\Delta \theta$, and averaging radially over the width of the ring, we obtain
the 1D value of the increment of orthoradial stress over 
$\Delta \theta$:
\begin{equation}
  \label{eq:Delta:stt1D}
 \Delta \sig_{\theta\theta}^{1\mathrm{D}} =  
\langle r \, \Tt \rangle_r \, \Delta \theta 
\end{equation}
This value is compared with the radially-averaged increment of orthoradial 
stress inferred by BISM $\langle \Delta \sig_{\theta\theta} \rangle_r$ 
(Figs.~\ref{fig:validation:exp}g-h). 
The excellent agreement found between experimental
($\langle r \, \Tt \rangle_r \, \Delta \theta$)
and inferred ($\langle \Delta \sig_{\theta\theta} \rangle_r$) 
1D stresses is quantified by a coefficient of determination
$R^2_{\mathrm{ring}}=0.99$, see Fig.~\ref{fig:validation:exp}h.
To check that BISM allows to infer absolute stress values,
we calculate the average pressure $\langle P_{\mathrm{exp}} \rangle$
from traction force data (see \emph{ST 1.4}):
\begin{equation}
  \label{eq:avg:pressure}
 \langle P_{\mathrm{exp}} \rangle = \frac{1}{2} \,\langle t_r r \rangle 
\end{equation}
where $ \langle \; \rangle$ denotes spatial averaging over the 
whole domain. We obtain 
$\langle P_{\mathrm{exp}} \rangle = - 2.18 \, \mathrm{kPa.\mu m}$,
in agreement with the average inferred pressure
$\langle P_{\mathrm{inf}} \rangle = - 2.17 \, \pm \, 0.94 \; 
\mathrm{kPa.\mu m}$.

We conclude that BISM is readily applicable to experimental traction force 
data, and has been validated on experimental data without 
reference to a  specific rheological model of the tissue.

\subsection{Comparison with monolayer stress microscopy} 
\label{sec:results:compare}

Monolayer stress microscopy, as introduced in \cite{Tambe2011}, 
assumes that the cell monolayer is a linear, isotropic elastic body.
Given traction force data, MSM consists in finding the stress field 
that minimizes an energy functional of the cell monolayer. MSM is thus 
straightforward to implement using \texttt{FreeFem++} \cite{hecht2012new},
a finite element software based on the same variational approach 
(see \emph{ST 1.1}).

A simpler implementation of MSM, that also assumes an elastic cell monolayer 
rheology, has been proposed recently \cite{Moussus2014}. We call this method 
MSM$u$ since it does not require the calculation of traction forces 
and uses the substrate displacement field $u$ as input data.
MSM$u$ further assumes that the displacement field $u$ is continuous 
at the interface between substrate and cells: tissue internal stresses 
are computed directly from substrate displacements. In practice,
we calculate substrate displacements from the traction force data set,
given numerical values
of the substrate elastic modulus $E_{\mathrm{sub}} = 5 \, \mathrm{kPa}$ 
and Poisson ratio $\nu_{2 \mathrm{D}\,\mathrm{sub}} = 0.5$ \cite{Moussus2014}. 

By analogy with MSM, we introduce a stress estimation method, named MSM$\eta$,
that assumes a viscous rheology for the cell monolayer.
Thanks to the variational formulation, we  compute the velocity 
field given the force traction field with \texttt{FreeFem++}, 
and estimate the stress field given numerical values of the
viscosity coefficients. Of note, other variants of MSM could be implemented
assuming other tissue rheologies consistent with a variational 
formulation \cite{Tlili2015}.

The comparison relies on three numerical simulations: 
in addition to the \emph{Viscous} and \emph{Elastic 1} data sets studied above,
another simulation of an elastic tissue, named \emph{Elastic 2}, has been 
performed, using the elastic coefficients 
$E=10 \, \mathrm{kPa \,\mu m}$ and $\nu_{2 \mathrm{D}}=0.5$, 
as advocated in \cite{Moussus2014}.
Each variant of MSM assumes a tissue rheology, and thus  relies
on a set of material parameters. 
To perform MSM and MSM$u$, we use the same tissue elastic 
coefficients as in the \emph{Elastic 1}  \cite{Tambe2011}
and \emph{Elastic 2} \cite{Moussus2014} simulations respectively.
To perform MSM$\eta$, we use the same tissue viscosity coefficients 
as in the \emph{Viscous} simulation.

Table~\ref{tab:MSM} summarizes our results and compares the accuracy
of the different methods (see also Fig.~S9 
in the Supporting Material for visual comparison). 
In all cases, the values of $R^2_{\sig}$ are closer to $1$ for BISM, which 
performs better than MSM, MSM$u$ and MSM$\eta$.
By construction, BISM seems less sensitive to experimental noise 
than deterministic stress microscopies.
For each variant of MSM, accuracy is maximal for the data set
generated according to the same rheological hypothesis,
\emph{i.e.} \emph{Elastic 1}, \emph{Elastic 2} and \emph{Viscous} for 
MSM, MSM$u$ and MSM$\eta$ respectively. Unsurprisingly, a  mismatch 
between numerical simulation and stress microscopy methods,
either in the values of material parameters or in the choice of a rheology, 
leads to a lower value of the coefficient of determination.

We also applied MSM to the experimental data set
studied in  \emph{Validation: experimental data}. 
Following the same protocol, we obtained a larger 
dispersion of inferred values than with BISM. 
The inferred data set obtained by MSM is characterized by a lower
coefficient of determination $R^2_{\mathrm{ring}}=0.32$,
far below the BISM value $R^2_{\mathrm{ring}}=0.99$.

All existing stress microscopies infer the stress field
up to an additive null vector $\sigma^0$ such that 
$\mathrm{div} \, \sigma^0=0$. Classically \cite{landau1975elasticity}, 
null vectors of the linear problem $\mathrm{div} \, \sigma= \vec{t}$
are related to the Airy stress function $\chi$ through 
$\sigma^0_{xx}=\frac{\partial^2\chi}{\partial y^2}$, 
$\sigma^0_{yy}=\frac{\partial^2\chi}{\partial x^2}$ and 
$\sigma^0_{xy}=-\frac{\partial^2\chi}{\partial x\partial y}$.
Since BISM infers faithfully the mean stress $\langle \sigma \rangle$
in confined geometries, it limits the class of undetectable stresses to 
\emph{zero-mean} stress fields that verify both $\mathrm{div} \, \sigma^0=0$
and the boundary conditions \eqref{eq:BC:general}.
Alternative methods will be necessary to ascertain the relevance
of these special solutions to cell monolayer mechanics.

 \begin{table}[!t]
   \centering
\begin{tabular}{ccccc}
  \hline Rheology
    & \hspace*{0.1cm} BISM \hspace*{0.1cm} 
    & \hspace*{0.1cm} MSM \hspace*{0.1cm} 
    & \hspace*{0.1cm} MSM$u$ \hspace*{0.1cm} 
    & \hspace*{0.1cm} MSM$\eta$ \hspace*{0.1cm} \\
  \hline
 \emph{Viscous}  & $\quad 0.96 \quad$ & $\quad -0.52 \quad$ 
& $\quad 0.48 \quad$ & $\quad 0.91 \quad$ \\
  \emph{Elastic 1}  & $\quad 0.97 \quad$ & $\quad 0.88\quad$ & $\quad 0.73\quad$
& $\quad 0.80 \quad$ \\
  \emph{Elastic 2}  & $\quad 0.99 \quad$ & $\quad 0.61\quad$ & $\quad 0.85\quad$
& $\quad 0.67 \quad$ \\
  \hline
\end{tabular}
\smallskip
\caption{\label{tab:MSM} 
\textbf{Comparison with monolayer stress microscopy.}
Coefficients of determination  $R^2_{\sig}$ obtained  with BISM, 
MSM, MSM$u$ and MSM$\eta$ (see text for definitions).
Traction force data sets were obtained with material parameter values
$\eta=10^{3} \, \mathrm{kPa \, \mu m\, s}$, $\eta'=\eta$ (\emph{Viscous});
$E=10^2 \, \mathrm{kPa \,\mu m}$, $\nu_{2 \mathrm{D}}=0.5$ \cite{Tambe2011}
(\emph{Elastic 1}), $E=10 \, \mathrm{kPa \,\mu m}$,
$\nu_{2 \mathrm{D}}=0.5$ \cite{Moussus2014} (\emph{Elastic 2}).
A white noise of  relative amplitude $5\%$ is added in all cases.
}
\end{table}

\section{Conclusion}
\label{sec:discussion}

Bayesian inversion stress microscopy  estimates the 
internal stress field of a cell monolayer given  traction force data. 
Validation on both numerical and experimental data shows that the 
method works reliably independently of the tissue 
rheology, of its geometry, and of the boundary conditions imposed on 
the stress field. As a consequence, BISM should apply equally well to 
isolated cell assemblies and to patches of cells within a larger
monolayer. Since the hypotheses made pertain to statistics 
(Bayesian inversion), we checked that the method is robust to changes 
in the underlying statistical model, in particular to changes in the prior.
Importantly, its statistical nature leads to a simple, natural definition 
of an error bar on the stress estimate. It is compatible with the 
level of experimental noise and with the spatial resolution
typical of traction force microscopy. Last, BISM is more accurate
than MSM, and its accuracy is less sensitive to the rheology of the tissue
than all variants of monolayer stress microscopy. 
BISM is quite general since it relies on the laws of mechanics
and on reasonable and robust statistical assumptions.
It  can therefore be applied to other active materials strongly 
interacting with a soft substrate, provided that the height of the system
is small compared to its planar spatial extension.

We analyzed traction force \emph{images}, \emph{i.e.} spatial data at a given, 
fixed time. However, BISM does not rely on an assumption of quasi-stationarity,
and would apply equally  well to spatio-temporal data, \emph{i.e.} to traction 
force \emph{movies}. Our preliminary results suggest that combining 
Bayesian inversion with Kalman filtering then further improves accuracy.

To date, the modeling of cell monolayer mechanics typically relies on a 
\emph{forward} approach: assumptions made on tissue rheology 
are validated indirectly through predictions made on (measurable) tissue 
kinematics. A reliable measurement of the internal stress field paves the way
to \emph{inverse} approaches, where the combination of stress 
with kinematic data, such as the strain rate field or the cell-neighbour 
exchange rate field, would allow to read out constitutive equations
from data, and to infer the values of material parameters.

\section*{Acknowledgements}
\label{sec:ack}

We would like to thank Cyprien Gay, Fran\c{c}ois Graner, Yohei Kondo and 
Sham Tlili for stimulating discussions.

Financial supports from the Human Frontier Science Program 
(grant RGP0040/2012), the European Research Council under the 
European Union's Seventh Framework Programme (FP7/2007-2013)/ERC 
grant agreement n$^o$ 617233, and the Mechanobiology Institute are 
gratefully acknowledged. S.J. acknowledges the Merlion-2014 programme
of the French Ministry of Foreign Affairs. B.L. acknowledges the 
Institut Universitaire de France.


\newpage
\section*{SUPPORTING TEXT}

\setcounter{section}{0}
\setcounter{figure}{0}
\setcounter{table}{0}
\setcounter{equation}{0}

\section{Computational aspects}
\label{sec:comput}

\subsection{Numerical simulations}
\label{sec:comput:freefem}

We use the finite element software \texttt{FreeFem++} \cite{hecht2012new} 
to solve numerically
the conservation equations and generate stress and traction force data
for several rheologies (viscous and elastic) and several geometries 
(square and disc).
\texttt{FreeFem++} requires the weak formulation of evolution equations.
The momentum conservation equation $\mathrm{div} \, \sig = \vec{t}$ reads
in variational form 
\[
 \int_{\Omega} {\vec \nabla} {\sig}.{\vec w}= \int_{\Omega}{\vec t}.{\vec w}
\]
where ${\vec w}$ is a vector test function and $\Omega$ is the domain 
of integration.
The boundary conditions $\sig \cdot \vec{n} = \vec{0}$,
where $\vec{n}$ denotes the vector normal to the edge, 
are implemented through an extra term integrated 
over the boundary $\delta\Omega$
\[
 \int_{\Omega} {\vec \nabla} {\sig}.{\vec w}= \int_{\Omega}{\vec t}.{\vec w}
+ \int_{\delta\Omega}{\sig \vec{n}}.{\vec w}
\]

\subsubsection*{Viscous liquid}
\label{sec:comput:compressible}

The constitutive equation of a viscous, compressible liquid
reads \cite{landau1987fluid}
\[
\sig=\eta \, \left( \vec{\nabla}\vec{v}+(\vec{\nabla}\vec{v})^t 
\right) +
\eta'  \, \left( \vec{\nabla}.\vec{v} \right) I
\]
where $\eta$ and $\eta'$ are the shear and compression coefficients of
viscosity. 
Taking into account friction with the substrate,
with a friction coefficient $\xi_v$, and a motility
force field $\vec{f}_{\mathrm{act}}$, the total traction force $\vec{t}$ reads
\[
\vec{t}=\xi_v\,\vec{v} - \vec{f}_{\mathrm{act}}
\]
The active force $\vec{f}_{\mathrm{act}}$ is generated by
$n_d$ force dipoles randomly distributed in space, see below. 

Using the momentum conservation equation, 
we obtain an equation on $\vec{v}$ 
\[
  -\eta \Delta \vec{v} - 
  (\eta+\eta') \, \vec{\nabla}(\vec{\nabla}.\vec{v}) + \xi_v \vec{v}
  = \vec{f}_{\mathrm{act}}
\]
whose weak form reads 
\[
\int_{\Omega}{\eta \vec \nabla} {\vec v}.{\vec \nabla} {\vec w} 
- \int_{\Omega}{ (\eta+\eta') \, \vec{\nabla}(\vec \nabla}. {\vec v}). {\vec w}+
\int_{\Omega}{\xi_v \vec v}.{\vec w} = 
\int_{\Omega}  {\vec f}_{\mathrm{act}}.{\vec w}
\]
We solve this equation for the velocity field with \texttt{FreeFem++} 
on a $100 \times 100 \, \mu\mathrm{m}^2$  square with 
boundary conditions $\sig \cdot \vec{n} = \vec{0}$,
and with material parameter values typical of cell monolayers:
$\xi_v=10^0 \,\mathrm{kPa\, \mu m}^{-1}\mathrm{s}$ \cite{Cochet-Escartin2014}, 
$\eta=10^{3} \, \mathrm{kPa \, \mu m\, s}$ \cite{Harris2012}, 
$\eta'= \eta$. To the best of our knowledge, the second coefficient of
viscosity $\eta'$ has not been measured in a tissue, but
theoretical reasoning suggests $\eta'= 2 \, \eta$ for a thin monolayer,
due to 3D incompressibility \cite{Jenkins1997}.
The stress and traction force fields are derived from the velocity field 
and sampled over a regular cartesian grid with $C=R=50$ and 
$N= C \times R = 2500$ (see Fig.~3a-b).

\subsubsection*{Elastic solid}
\label{sec:comput:elastic}

In the limit of small deformations, 
the constitutive equation of an elastic body reads
\cite{landau1975elasticity}
\[
\sig=\mu \, \left( \vec{\nabla}\vec{u}+(\vec{\nabla}\vec{u})^t \right) 
+\lambda\, ( \vec{\nabla} . \vec{u} )  \, I
\]
with Lam\'e coefficients $\lambda, \mu$.
With a (solid) friction force proportional to the displacement,
the traction force reads 
$\vec{t}=\xi_u \vec{u}-\vec{f_{act}}$,
and the momentum conservation equation
\[
  -\mu \Delta \vec{u}
  - (\mu+\lambda) \, \vec{\nabla}(\vec{\nabla}.\vec{u}) + \xi_u \vec{u}
  = \vec{f}_{\mathrm{act}}
\]
is formally identical to that obtained for the flow of a viscous liquid,
substituting $\vec{v}$ by $\vec{u}$ and  the material parameters
$\eta$, $\eta'$, $\xi_v$ by $\mu$, $\lambda$, $\xi_u$.
The numerical resolution is therefore similar to that described above
for a viscous liquid. Parameter values differ: for instance a 
realistic Poisson ratio $\nu_{2\mathrm{D}}=\frac{\lambda}{\lambda+2\mu}=0.5$ 
imposes $\lambda=2\mu$.

\subsubsection*{Active forces}
\label{sec:comput:fact}

The active force $\vec{f}_{\mathrm{act}}$ is generated by
$n_d$ force dipoles defined in tensor form as
\[
p^n(\vec{x})=\bigl[p^n_{\textrm{tr}}  \, 
\bigl(\begin{smallmatrix}
1 & 0\\0 & 1
 \end{smallmatrix} \bigr) 
- p^n_{\textrm{dev}} \, 
(\begin{smallmatrix}
 cos 2\theta_n & sin 2\theta_n\\sin 2\theta_n & -cos 2\theta_n
  \end{smallmatrix}\bigr)
\bigr] \, \delta(\vec{x}-\vec{x}_n), \quad n=1,\ldots,n_d
\]
with positions $\vec{x}_n$, amplitudes $p^n_{\textrm{tr}}$ and 
$p^n_{\textrm{dev}}$ for the trace and deviator, and orientations $\theta_n$. 
The dipole positions $\vec{x}_n$ and orientations $\theta_n$ are 
random variables uniformly distributed 
over the spatial domain $\Omega$ and $[0, 2\pi]$ respectively.
In practice, the force dipoles are implemented 
using a finite-size Gaussian approximation of the delta function,
with a spatial extension $2d = 10 \, \mu$m of the order of a typical 
cell size. The components of the active force read
\[
 f_{\mathrm{act},i}(\vec{x}) =\sum_{n=1}^{n_d}\partial_j p^n_{ij}(\vec{x}) 
\]
with  $n_d=100$ force dipoles of typical amplitude $10^{-2} \,\mathrm{kPa}$ 
\cite{du2005force}.
Since the amplitude of traction forces is known to be larger
close to the edge than in the bulk, 
the coefficients $p^n_{\textrm{tr}}$ and $p^n_{\textrm{dev}}$ are set
proportional to $1+r/l_p$, where $r$ is the distance to the center
of the domain and $l_p = 5 \, \mathrm{\mu m}$ is a penetration length  
\cite{du2005force}.

\subsubsection*{MSM implementation}
\label{sec:comput:msm}

Monolayer stress microscopy \cite{Tambe2011} consists in finding,
given traction force data, the displacement field ${\vec u}$ that verifies 
the force balance equation obtained for a purely elastic stress.
Thanks to the hypothesis made on tissue rheology, the problem 
becomes well-posed since two components of the displacement vector 
are deduced from two components of the traction force vector. 
Once  ${\vec u}$ is known, the calculation of the stress 
depends on the numerical values of tissue elastic coefficients.
We implement MSM with \texttt{FreeFem++}, solving the weak form
\[
\int_{\Omega}{\mu \vec \nabla} {\vec u}.{\vec \nabla} {\vec w} - 
\int_{\Omega}{(\mu+\lambda) \vec{\nabla}(\vec \nabla}. {\vec u}). {\vec w}+
\int_{\Omega}{\vec t}.{\vec w} + 
\int_{\partial \Omega} \left({\sigma \vec n}\right).{\vec w} = 0
\]
for appropriate test functions $\vec{w}$,
with numerical or experimental data sets of traction forces.

The variant MSM$\eta$ assumes that the tissue has the rheology of
a viscous liquid. It is well-posed for the same reason, and consists 
in solving the weak form of the ``viscous'' force balance equation
\[
\int_{\Omega}{\eta \vec \nabla} {\vec v}.{\vec \nabla} {\vec w} - 
\int_{\Omega}{(\eta+\eta') \, \vec{\nabla}(\vec \nabla}. {\vec v}). {\vec w}+
\int_{\Omega}{\vec t}.{\vec w} + 
\int_{\partial \Omega} \left({\sigma \vec n}\right).{\vec w} = 0
\]
for the velocity field $\vec{v}$, from which the stress is computed 
given numerical values of the viscosities $\eta$ and $\eta'$.

\subsection{Polar coordinates}
\label{sec:comput:polar}

For cell monolayers confined within a ring, or within a disk, 
we use the polar coordinate system $(r,\theta)$.
The force balance equations for the stress tensor field
$\sig = \bigl(\begin{smallmatrix}
\srr & \str \\
   \srt & \stt
\end{smallmatrix} \bigr)$ 
and the traction force field
$\vec{T} = \bigl(\begin{smallmatrix}
\Tr\\
   \Tt
\end{smallmatrix} \bigr)$ 
read
\begin{eqnarray*}
\label{eq:force:r}
\frac{1}{r}\DP{(r\srr)}{r}+\frac{1}{r}\DP{\srt}{\theta}-\frac{\stt}{r} 
&=& \Tr \\
\label{eq:force:t}
\frac{1}{r}\DP{\stt}{\theta}+\frac{1}{r}\DP{(r\str)}{r}+\frac{\srt}{r} 
&=& \Tt
\end{eqnarray*}
where, as for the cartesian coordinate system, we do not enforce the 
symmetry of the stress tensor.

When discretizing the derivatives with respect to $r$ and $\theta$ 
to lowest order, the (cartesian) expression (8) for the matrix $A$ 
is no longer valid and must be replaced by
\[
  A_{(r,\theta)} = 
   \bigl(\begin{smallmatrix}
   A_r/r & -I/r & A_{\theta}/r & 0\\
    0& A_{\theta}/r & I/r & A_r/r
   \end{smallmatrix} \bigr)
\]
where $A_r$ and $A_{\theta}$ are sub-matrices that correspond to the 
discretized forms of the partial derivatives 
$\DP{}{r}  r$ and  $\DP{}{\theta}$ respectively.

\subsection{Grids}
\label{sec:comput:grids}

\subsubsection*{Numerical data}
\label{sec:comput:grids:num}

In order to implement the mechanics of a rectangular piece of tissue 
on a grid $G_i$ with $C$ columns for the $x$ coordinate and $R$ rows 
for the $y$ coordinate of traction force data, in reality we need to
simulate a grid $G_s$  with $2R+1$ rows and $2C+1$ columns,
with two traction force components and four stress components at each point
(see Fig.~1b). We extract from the odd-numbered columns and even-numbered rows 
the $\sig_{xx}$ and $\sig_{yx}$ components ($(C+1) \times R$ values), 
from the even-numbered columns and odd-numbered rows the $\sig_{yy}$ 
and $\sig_{xy}$ components ($C \times (R+1)$ values) and finally from 
the even-numbered columns and even-numbered rows the $t_x$ and $t_y$ 
values ($C\times R$ values).
When plotting the stress field, each stress component is interpolated
(second order interpolation in $\dx$) at a position at the center 
of each grid cell to obtain the required $C\times R$ values.
We use the same method in polar coordinates, 
replacing $x$ by $\theta$ and $y$ by $r$.

\subsubsection*{Experimental data}
\label{sec:comput:grids:exp}

Irrespective of the method, each experimental traction force data point
is associated with a given cell of the 
measurement grid $G_m$. We need to interpolate these experimental values 
on a regular grid $G_i$, either cartesian or polar depending on the geometry. 
For each cell of the numerical grid $G_i$, we first calculate the areas
of intersection with cells of the initial grid $G_m$, and compute an 
average, weighted by these areas, over the traction forces in the cells 
of $G_m$ with a non-zero overlap. This procedure yields the traction force 
in the cell of $G_i$, which we use for stress inference.

\subsection{Average stress from traction force data}
\label{sec:comput:mean}

Spatially averaged values of stress components can be 
calculated directly from traction force data assuming the boundary
condition $\sigma_{ij} \, n_j=0$. For completeness sake,
we adapt here the classical derivation \cite{landau1975elasticity} 
to the calculation of the average pressure in polar coordinates.

We use cartesian coordinates $(x_1, x_2) = (x,y)$, and
summation over repeated indices is assumed throughout.
Multiplying both terms of the force balance equation
\begin{equation}
  \label{eq:force:balance}
  \DP{\sigma_{ij}}{x_j} = t_i
\end{equation}
by $x_k$, and integrating over the domain $\Omega$, we obtain
\begin{align*}
  \int_{\Omega} t_i \, x_k  & = \int_{\Omega} \DP{\sigma_{ij}}{x_j} \, x_k\\
& =  \int_{\Omega} \DP{}{x_j}\left( \sigma_{ij} x_k \right)  -
\int_{\Omega} \sigma_{ij} \DP{x_k}{x_j}  \\
& = \int_{\partial \Omega}  \sigma_{ij} x_k \, n_j
- \int_{\Omega} \sigma_{ik} \\
& = 0 - \int_{\Omega} \sigma_{ik} 
\end{align*}
where we used $\DP{x_k}{x_j} = \delta_{kj}$ and $\sigma_{ij} \, n_j=0$.
Denoting as usual spatial averages by brackets $\langle \ldots \rangle$,
we find \cite{landau1975elasticity}:
\begin{equation}
  \label{eq:avg:stress}
  \langle \sigma_{ik}  \rangle =  - \langle t_i \, x_k \rangle \,.
\end{equation}
In the case of the numerical simulation leading to Fig.~3,
we checked that the average inferred stress values 
$  \langle \sigma_{ik}^{\mathrm{inf}}  \rangle$ agree with
the values $\langle \sigma_{ik}^{\mathrm{num}}  \rangle$
computed directly from the numerical traction force data
using \eqref{eq:avg:stress}. 

We next calculate the average pressure, invariant under a change
of coordinate system since $P = -\frac{1}{2}\mathrm{Tr} \sigma$:
\begin{align*}
  \langle P \rangle & 
= -\frac{1}{2}( \langle \sxx \rangle + \langle \syy \rangle ) \\
&=   \frac{1}{2}( \langle t_x \, x \rangle + \langle t_y \, y \rangle ) 
\end{align*}
Switching to polar coordinates, we obtain
\begin{equation}
  \label{eq:avg:pressure}
  \langle P \rangle = \frac{1}{2}\langle t_r r \rangle \,.
\end{equation}

\section{Robustness}
\label{sec:robust}

We describe here the variations performed on BISM
in order to test its robustness, by varying
\emph{one by one} each feature of the statistical model
(Secs.~\ref{sec:robust:stat}), and of the numerical simulation 
(Sec.~\ref{sec:robust:sim}). Precise definitions and implementations 
of the statistical models are given in Sec.~\ref{sec:robust:models}.

\subsection{Robustness to variations of the statistical model}
\label{sec:robust:stat}

Variations made on the prior are the following.

\begin{itemize}
\item[\emph{(i)}]{Distribution of the stress
\begin{itemize}
\item[\emph{P1}]{Laplace prior: 
amounts to regularization with a L$_1$ norm \cite{tarantola2005inverse}.
}
\item[\emph{P2}]{Smoothness prior: penalizes spatial variations 
of the stress field \cite{kaipio2006statistical}.}
\item[\emph{P3}]{Unequal standard deviations 
$s_0^{\mathrm{tr}} \neq s_0^{\mathrm{dev}}$ 
for the trace and deviator of the stress.}
\item[\emph{P4}]{Non-zero mean value of the stress: 
$\sig_0 = \bigl(\begin{smallmatrix}
\sigma_{0 xx} & \sigma_{0 yx} \\
   \sigma_{0 xy} & \sigma_{0 yy}
\end{smallmatrix} \bigr)$.}
\end{itemize}
}
\item[\emph{(ii)}]{Equality of the shear components
\begin{itemize}
\item[\emph{P5}]{
Not enforced: $\axy = 0$ in Eq.~(10).} 
\end{itemize}
}
\item[\emph{(iii)}]{Boundary conditions
\begin{itemize}
\item[\emph{P6}]{
Not enforced: $\aBC = 0$ in Eq.~(10). 
}
\end{itemize}
}
\end{itemize}

Further, we implement the following variations.
\begin{itemize}
\item[]{
\begin{itemize}
\item[\emph{S1}]{
Inverse-Gamma hyperprior distribution: a weakly informative, but proper 
hyperprior for $s^2$ and $s_0^2$ \cite{gelman2014bayesian}.
}
\item[\emph{S2}]{
Student likelihood distribution: admits tails fatter than Gaussian.}
\item[\emph{S3}]{Expectation-Maximization algorithm: 
optimizes a likelihood function that depends on 
"latent" variables, here $s^2$ and $s_0^2$ \cite{McLachlan2008}.
}
\end{itemize}
}
\end{itemize}

Table~S\ref{tab:robust:stat} lists the values of $R^2_{\sig}$ 
thus obtained, given the same traction force and stress data sets as for BISM. 
Values of $1-R^2_{T}$ and of $\chi^2_T$ were always very close to $0$ 
and are therefore omitted.

\begin{table*}[!t]
   \centering
\begin{adjustwidth}{-0.6in}{0in} 
\begin{tabular}{ccccccccccc}
  \hline
      & BISM & \emph{P1} & \emph{P2} & \emph{P3} & \emph{P4} & \emph{P5}
& \emph{P6} & \emph{S1} & \emph{S2} & \emph{S3} \\       
  \hline
    $\quad R^2_{\sig} \quad$  & $\quad 0.96 \quad$ & $\quad 0.95 \quad$
& $\quad 0.96 \quad$ & $\quad 0.96 \quad$ & $\quad 0.95\quad$ & 
$\quad 0.75 \quad$ & $\quad 0.53 \quad$ & $\quad  0.96 \quad$ & 
$\quad 0.96 \quad$ & $\quad 0.94 \quad$\\ 
  \hline
\end{tabular}
\captionsetup{labelformat=supptable}
\caption{
\label{tab:robust:stat} 
\textbf{Robustness to variations of the statistical model.}
}
\end{adjustwidth}
\end{table*}

\begin{table*}[!t]
   \centering
\begin{adjustwidth}{-0.6in}{0in} 
\begin{tabular}{ccccccccccc}
  \hline
      & BISM & \emph{N1}$_-$ & \emph{N1}$_+$ & \emph{N2} & \emph{N3} 
& \emph{N4} & \emph{N5} & \emph{N6}    \\       
  \hline
    $\quad R^2_{\sig} \quad$  & $\quad 0.96 \quad$ & $\quad 0.99 \quad$
& $\quad 0.83 \quad$ & $\quad 0.97 \quad$ & $\quad 0.91 \quad$ 
& $\quad 0.85 \quad$ & $\quad 0.93\quad$ & $\quad 0.96 \quad$  \\ 
  \hline
\end{tabular}
\captionsetup{labelformat=supptable}
\caption{
\label{tab:robust:sim} 
\textbf{Robustness to variations of the numerical simulation.}
}
\end{adjustwidth}
\end{table*}

\subsection{Robustness to variations of the numerical simulation}
\label{sec:robust:sim}

Variations made on the numerical simulations are the following.

\begin{itemize}
\item[\emph{N1}]{Material parameters: keeping 
$\xi=10^0 \,\mathrm{kPa\, \mu m}^{-1}\mathrm{s}$ and 
$\eta=10^3 \,\mathrm{kPa\, \mu m}\,\mathrm{s}$, we vary $\eta'$:
$\eta' = 10^{-1} \, \eta$ (\emph{N1}$_-$);
$\eta' = 10^{1} \, \eta$ (\emph{N1}$_+$).
}
\item[\emph{N2}]{Elastic rheology: with a Young modulus 
$E=10^2 \, \mathrm{kPa\,\mu m}$ \cite{Harris2012,Delarue2014}, 
a Poisson ratio $\nu_{2 \mathrm{D}}=0.5$ 
\cite{Tambe2013}) and a (solid) friction coefficient 
$\xi_u=1 \, \mathrm{kPa\;\mu m^{-1}}$ \cite{banerjee2012contractile}.
}
\item[\emph{N3}]{
No-slip boundary conditions: $\vec{v}=\vec{0}$ at the edge instead of 
$\sig \cdot \vec{n} = \vec{0}$, with $\aBC = 0$.
}
\item[\emph{N4}]{Subsystem of a larger tissue: central part of size 
$R \times C$ of a system of size $3R \times 3C$, with $\aBC = 0$, since the 
boundary conditions on the subsystem are unknown.}
\item[\emph{N5}]{Circular geometry: see Fig.~S\ref{fig:disc}}
\item[\emph{N6}]{Student distribution function of the measurement noise:
$St(\nu=5)$ with unchanged variance $s_{\mathrm{exp}}^2$. 
}
\end{itemize}
Table~S\ref{tab:robust:sim} lists the values of $R^2_{\sig}$ obtained
by applying BISM to the traction force field obtained from each
of these numerical simulations. We also study the influence of 
spatial resolution and of the amplitude of measurement noise on the
accuracy of inference, see Fig.~4.

\begin{figure}[!t]
\centering 
\showfigures{
(a)\includegraphics[scale=0.3]{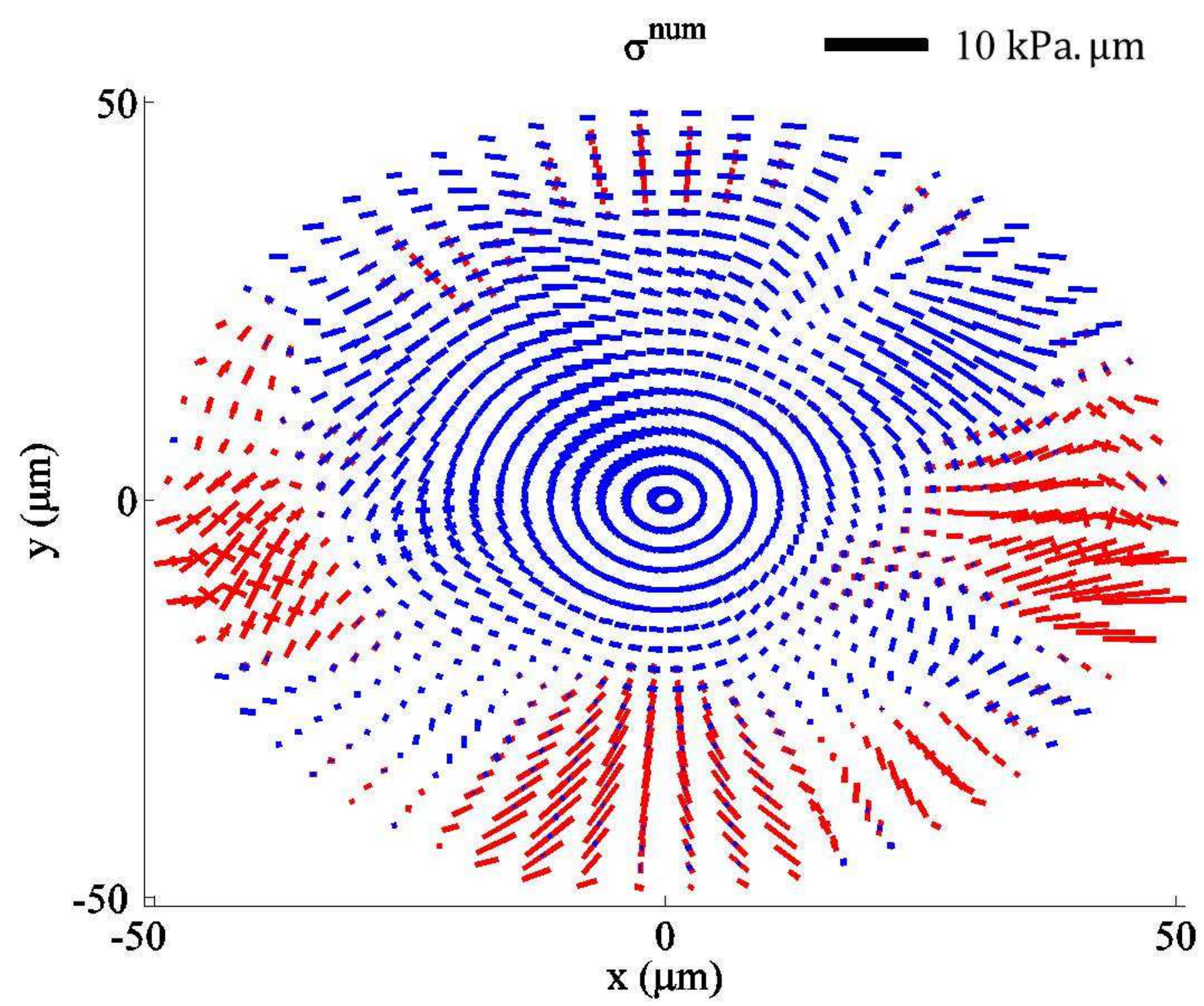}
\hspace*{1cm}
(b)\includegraphics[scale=0.3]{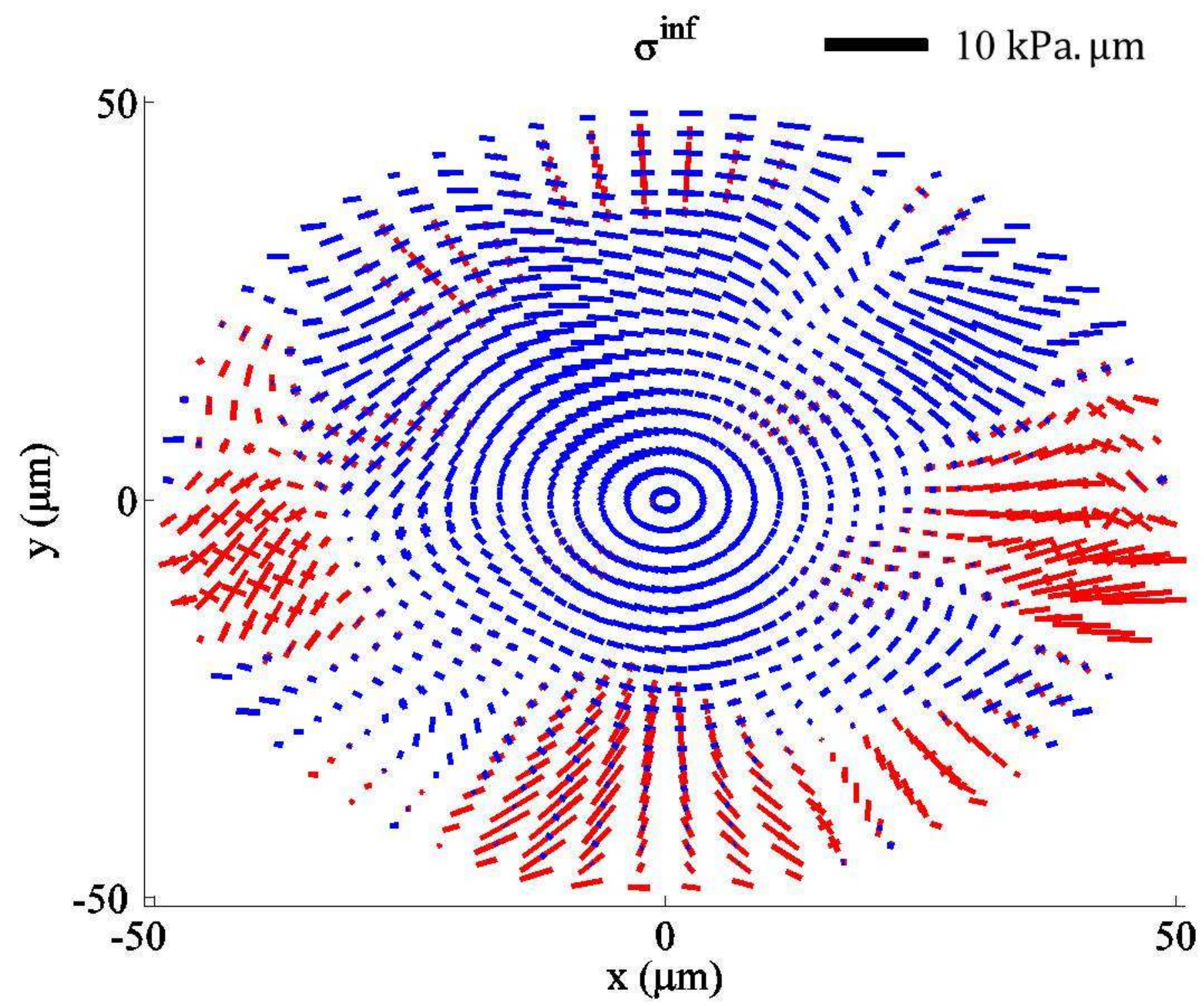}
}
\captionsetup{labelformat=suppfig}
\caption{\label{fig:disc}
\textbf{Polar coordinates.}
Monolayer confined to a disc (variation \emph{N5}). 
Excellent agreement is found between (a) simulated stress $\sig^{\mathrm{num}}$
and (b) inferred stress $\sig^{\mathrm{inf}}$.
The stress fields are computed and represented as in 
Fig.~3, on a $20 \times 60$ polar grid in a disc of radius $50\,\mu$m.
}
\end{figure}

\subsection{Statistical models}
\label{sec:robust:models}

\subsubsection{Laplace prior distribution ({\it P1})}
\label{sec:robust:models:pi}

Since MAP estimation is untractable analytically with a Laplace prior, 
we use a hierarchical construction \cite{eltoft2006multivariate}. 
A multivariate Laplace distribution of parameter $\lambda$
and covariance matrix $2B/\lambda$ is obtained by
compounding a Gaussian distribution 
$\mathcal{N}(\vec{\sig} \mid \vec{0}, s_0^2 B)$
with an exponential hyperprior for $s_0^2$:
$H(s_0^2 \mid \lambda)=\frac{\lambda}{2} \, 
\exp\left(-\frac{\lambda s_0^2}{2}\right)$,
and marginalizing over $s_0^2$:
\[
\pi_{\mathrm{L}}(\vec{\sig})= \int 
\mathcal{N}(\vec{\sig} \mid \vec{0}, s_0^2 B)
\, H(s_0^2 \mid \lambda) \, \mathrm{d}s_0^2
\] 
Tensor symmetry and boundary conditions are taken into account
as above, see Fig.~S\ref{fig:hierarchical:var}a. 
Using Jeffreys' hyperprior for 
$\lambda$, $H(\lambda) \propto \frac{1}{\lambda}$,
MAP estimation leads to the iteration rule
\begin{eqnarray*}
{s^2}_{(k)} &=&\frac{1}{2N+2} \; \| \vec{T} - A \vec{{\sig}}_{\Pi \,(k)} \|^2
\\
\label{eq:hyperparameters:laplace:s0}
\lambda_{(k)} & = & \frac{2\lambda_{(k-1)}}{4N+2(C+R)} 
\left(\sqrt{1+ \frac{4 \lambda_{(k-1)}}{(4N+2(C+R))^2}
\, \vec{\sig}_{\Pi (k)}^t \, B^{-1} \, \vec{\sig}_{\Pi (k)}} -1 \right)^{-1}
\end{eqnarray*}
The regularization parameter is defined as
\[
\Lambda_{\mathrm{L}} = \frac{1}{2}  \, \lambda \,l^2 \,s^2
\]

\subsubsection{Smoothness prior distribution  ({\it P2})}
\label{sec:robust:models:cov}

We naturally expect the tissue stress field  to be a smooth function of space.
This information can be embedded in the prior distribution function
\cite{dembo1996imaging,kaipio2006statistical},
by penalizing large spatial gradients: (variation \emph{P2}):
\[
\pi_{\mathrm{s}}(\vec{\sig}) \propto 
\mathrm{exp} \left[ -\frac{  \dx^2 \| \vec{ \nabla \sig} \|^2 +
\axy^2 \| \vec{\sigma}_{xy} -\vec{\sigma}_{yx} \|^2 + 
\aBC^2 \| \vec{\sig}_{BC} \|^2 }{2s_0^2} \right]
\]
Discretizing the gradient operator to first order in $\dx$ leads to
a modified effective covariance matrix $B_{\mathrm{s}}$ with
$\pi_{\mathrm{s}}(\vec{\sig})=
\mathcal{N}(\vec{\sig} \mid \vec{0}, S_{0\,\mathrm{s}} = s_0^2 \, B_{\mathrm{s}})$.
The iteration rule (19-20) is unchanged.

\subsubsection{Inverse-Gamma hyperprior distribution  ({\it S1})}
\label{sec:robust:models:H}

A natural alternative to the non-informative, improper Jeffreys' distribution
for the hyperprior is the weakly informative, proper
inverse-Gamma distribution
\[
H(s^2 \mid \epsilon, s^2_H)= IG(\epsilon,\epsilon s^2_H) 
= \frac{(\epsilon s^2_H)^\epsilon}{\Gamma(\epsilon)} \,
\left(\frac{1}{s^2}\right)^{\epsilon+1}
\, \mathrm{exp} \left[ -\frac{\epsilon s^2_H}{s^2} \right]
\]
It tends to Jeffreys' distribution when the parameter
$\epsilon$ goes to zero, and, conveniently, is conjugate 
to the Gaussian distribution \cite{gelman2014bayesian}.
Its mode is equal to $\frac{\epsilon}{\epsilon+1}s^2_H$. 

For simplicity, we use the same inverse-Gamma hyperprior distribution for 
$s^2$ and $s_0^2$, and set the additional parameters to constant values
$\epsilon=10^{-1}$, $s^2_H=1 \, \mathrm{kPa^2}$ and 
$s^2_{0 H}=1 \,\mathrm{kPa^2 \,\mu m^2}$. 
The evolution equations read (compare with Eqs.~(19-20))
\begin{eqnarray*}
\label{eq:hyperparameters2:s}
s^2_{(k)} &=&
\frac{1}{2N+2+2\epsilon} \, 
\left( \| \vec{T} - A \vec{\sig}_{\Pi \,(k)} \|^2+2\epsilon s^2_H \right)
\\
\label{eq:hyperparameters2:s0}
s_{0(k)}^2&=& \frac{1}{4N+2(R+C)+2+2\epsilon} \,
\left( \vec{\sig}_{\Pi \,(k)}^t \, B^{-1} \, \vec{\sig}_{\Pi \,(k)} +2\epsilon s^2_{0 H} \right)
\end{eqnarray*}
Varying $\epsilon$, the accuracy of inference is excellent  as long as
$\epsilon < \epsilon_{\mathrm{max}} \approx 3\;10^{-1}$.
Above this value, the bias brought by the finite mean value of $s^2$
and $s_0^2$ compromises the inference.

\begin{figure*}
\centering 
\showfigures{
(a)\includegraphics[scale=0.35]{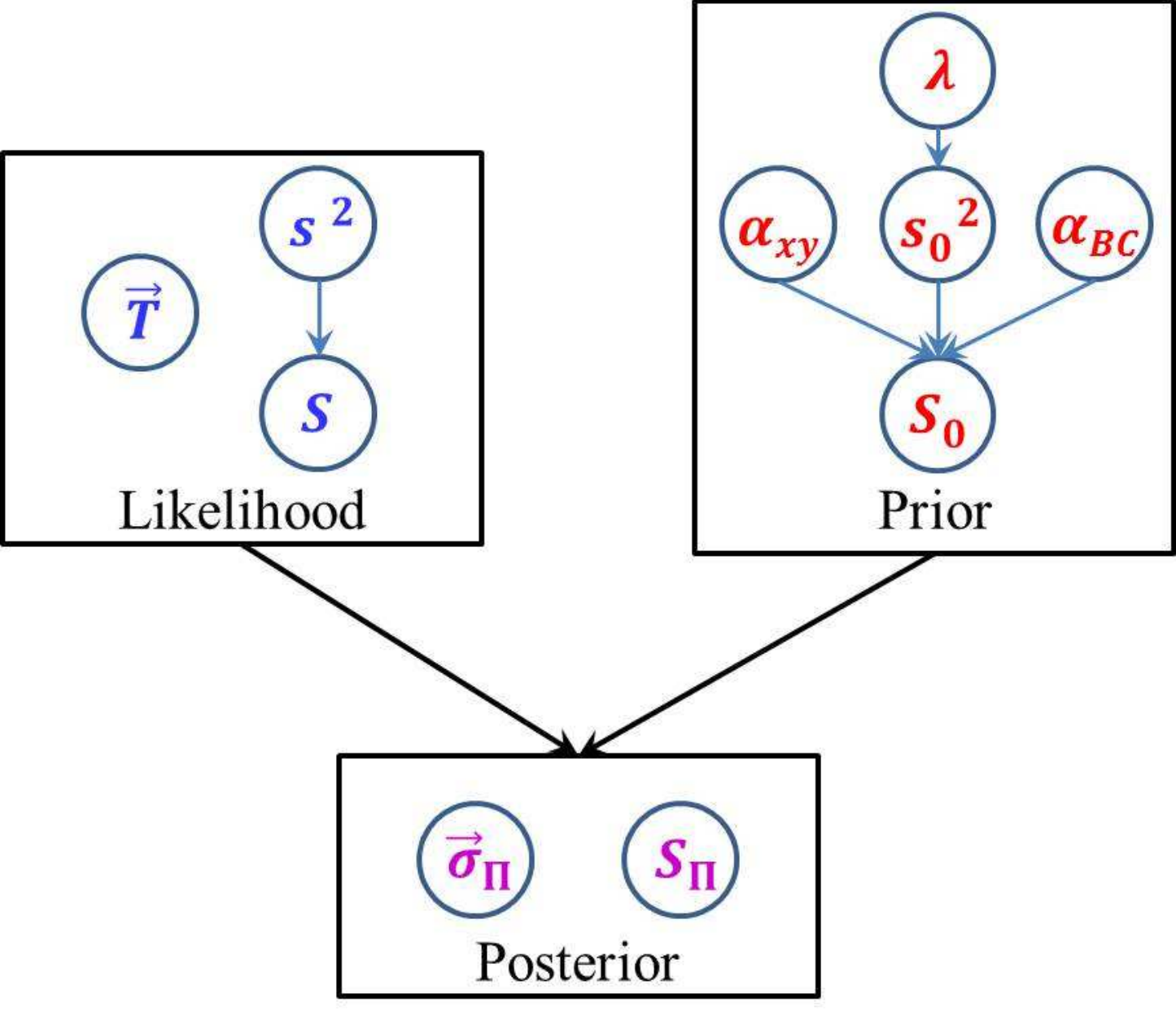}
\hspace*{1cm}
(b)\includegraphics[scale=0.35]{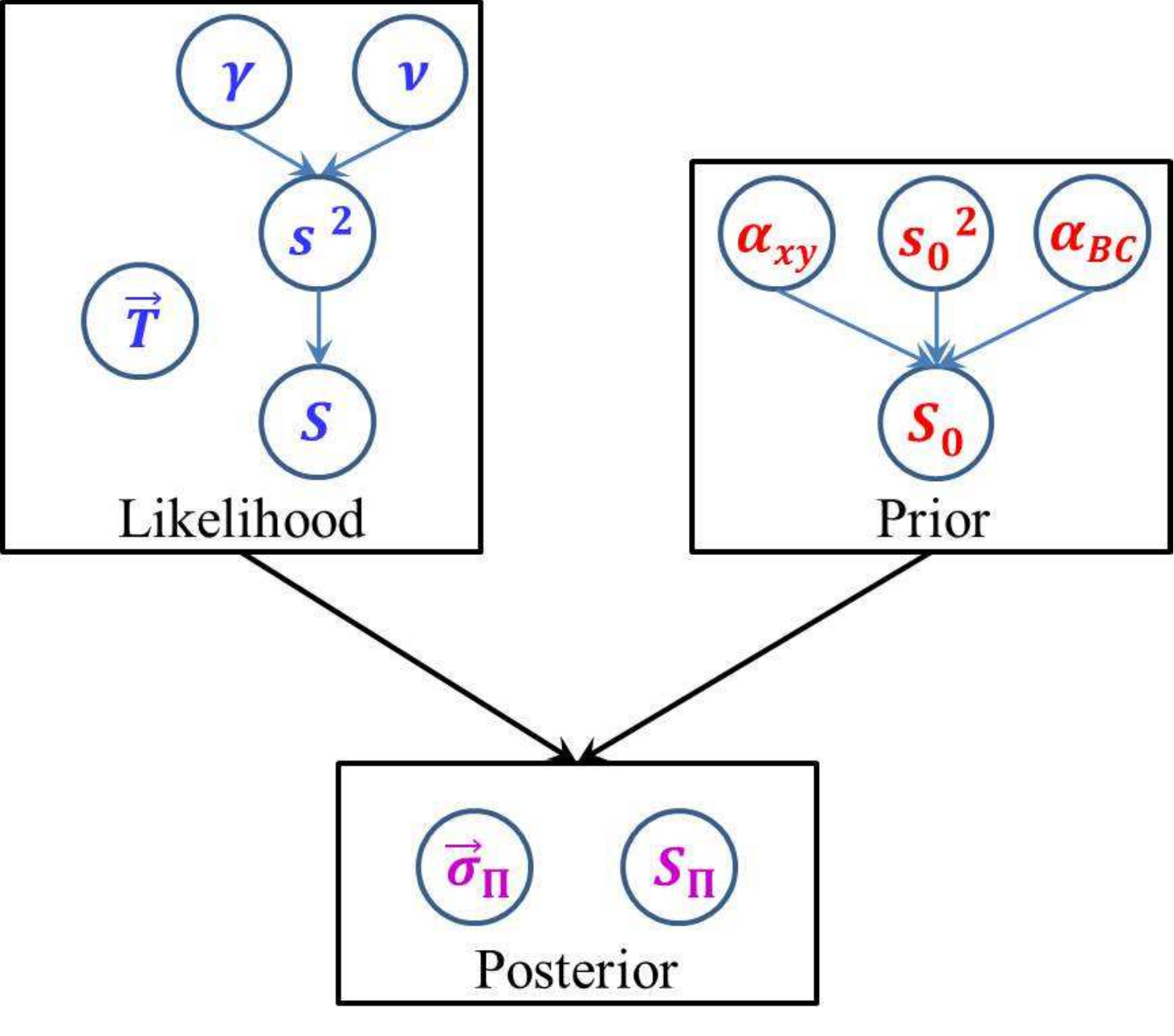}
}
\captionsetup{labelformat=suppfig}
\caption{\label{fig:hierarchical:var}
\textbf{Schematics of hierarchical Bayesian inversion.}
(a) Laplace prior (\emph{P1}).
(b) Student likelihood (\emph{S2}).
}
\end{figure*}

\subsubsection{Student likelihood distribution  ({\it S2})}
\label{sec:robust:models:L}

A simple choice of a zero-mean distribution with tails fatter than Gaussian 
is the non standardized Student distribution $St(\nu,0,\gamma)$.
It admits a Gaussian limit $St(\nu,0,\gamma) \to \mathcal{N}(0,\gamma)$ 
as $\nu \to \infty$, and has a finite variance 
$\frac{\nu \gamma}{\nu-2}$ if $\nu>2$.  

To implement a Student likelihood function in a tractable way, 
we use again a hierarchical approach 
\cite{beal2003variational} (see Fig.~S\ref{fig:hierarchical:var}b).
The Student distribution $St(\nu,0,\gamma)$ is obtained by
compounding a Gaussian distribution 
$\mathcal{N}(\vec{T} \mid A \vec{\sig}, s^2 \, I)$ 
with an inverse-Gamma distribution 
$H(s^2 \mid \gamma, \nu) = IG\left(\frac{\nu}{2},\frac{\nu\gamma}{2}\right)$
for the variance $s^2$, with Jeffreys' hyperpriors 
$H(\gamma) \propto \frac{1}{\gamma}$, $H(\nu) \propto 1$ for $\gamma$ 
and $\nu$, and by integrating over $s^2$
\[
L_{\mathrm{S}}(\vec{T} \mid \vec{\sig})= \int 
\mathcal{N}(\vec{T} \mid A \vec{\sig}, s^2 \, I) \,
H(s^2 \mid \gamma, \nu) \, H(\gamma) \, H(\nu) \,
\mathrm{d}s^2
\]

The iteration rule reads
\begin{eqnarray*}
\label{eq:hyperparameters:student:s0}
{s_0^2}_{(k)}&=& \frac{1}{4N+2(R+C)+2} \;
\vec{{\sig}}_{\Pi \,(k)}^t \, B^{-1} \, \vec{{\sig}}_{\Pi \,(k)}\\
\label{eq:hyperparameters:student:s}
\gamma_{(k)}&=& \frac{\nu_{(k-1)}-2}{\nu_{(k-1)}} \, \frac{1}{2N+\nu_{(k-1)}+2} 
\left( \| \vec{T} - A \vec{\sig}_{\Pi \,(k)} \|^2 + 
\nu_{(k-1)} \gamma_{(k-1)} \right)\\
\label{eq:hyperparameters:student:nu}
\psi\left(\frac{\nu_{(k)}}{2}\right) - \log\left(\frac{\nu_{(k)}}{2}\right)
 &=& \log\left(\frac{\nu_{(k-1)}-2}{\nu_{(k-1)}}\right) - 
\frac{\nu_{(k-1)}-2}{\nu_{(k-1)}} +1  
\end{eqnarray*}
where $\psi$ is the digamma  function, and the last equation is solved for
$\nu_{(k)}$. 
The regularization parameter is defined as
\[
\Lambda_{\mathrm{S}} = \frac{l^2 \, \gamma}{s^2_0}  \frac{\nu}{\nu-2}   
\]

\subsubsection{Expectation-maximization algorithm  ({\it S3})}
\label{sec:robust:models:hyperparameters}

A classical alternative to MAP optimization is the expectation-maximization 
(EM) algorithm \cite{McLachlan2008}. Once $\vec{\sig}_{\Pi \,(k)}$ is computed 
using the $(k-1)$th parameter values $s^2_{(k-1)}$ and $s_{0 (k-1)}^2$, we 
need to
\begin{itemize}
\item[\emph{E step}:]{define a function 
$Q$ as the expectation value over $\vec{\sig}$ of the log-posterior 
\[
Q(s^2, s_0^2 \mid s^2_{(k-1)}, s_{0 (k-1)}^2, \vec{\sig}_{\Pi \,(k)} ) = 
E_{\vec{\sig}} \left[ \log 
\Pi \left( \vec{\sig}, s^2, s_0^2 \mid 
\vec{T}, s^2_{(k-1)}, s_{0 (k-1)}^2, \vec{\sig}_{\Pi \,(k)} 
\right) \right]
\]
} 
\item[\emph{M step}:]{optimize this function Q over respectively
 $s^2$ and $s_0^2$ to calculate the $k$th parameter values 
$s^2_{(k)}$ and $s_{0 (k)}^2$
\begin{eqnarray*}
s^2_{(k)}&=& \mathrm{argmax}_{s^2} \; 
Q(s^2, s_0^2 \mid s^2_{(k-1)}, s_{0 (k-1)}^2, \vec{\sig}_{\Pi \,(k)} ) \\
s_{0 (k)}^2&=& \mathrm{argmax}_{s_0^2} \; 
Q(s^2, s_0^2 \mid s^2_{(k-1)}, s_{0 (k-1)}^2, \vec{\sig}_{\Pi \,(k)} ) 
\end{eqnarray*}
}
\end{itemize}
We obtain the iteration rule  \cite{tipping2001sparse} 
(compare with Eqs.~(19-20))
\begin{eqnarray*}
\label{eq:EM:hyperparameters:s}
s^2_{(k)} &=&\frac{1}{2N+2} \,
\left(
\| \vec{T} - A \vec{\sig}_{\Pi \,(k)} \|^2 
+ s^2_{(k-1)}
\left( 4N+2(R+C)- \frac{\mathrm{tr}\,S_{\Pi}}{s_{0(k-1)}^2} \right) 
\right)
\\
\label{eq:EM:hyperparameters:s0}
s_{0 (k)}^2&=& \frac{1}{4N+2(R+C)+2} \,
\left(
\vec{\sig}_{\Pi \,(k)}^t \, B^{-1} \, \vec{\sig}_{\Pi \,(k)}+\mathrm{tr}\;S_{\Pi}
\right) 
\end{eqnarray*}
Convergence is fast, similar to MAP estimation, but the calculation 
is more demanding computationally due to the need to evaluate 
$S_{\Pi}$ at each step.

\section{Parameter and hyperparameter optimization}
\label{sec:hyper:opt}

\subsection{Convergence}
\label{sec:opt:cv}

\begin{figure}[!t]
\centering 
\begin{adjustwidth}{-0.6in}{-0.65in} 
\showfigures{
(a)\includegraphics[scale=0.35]{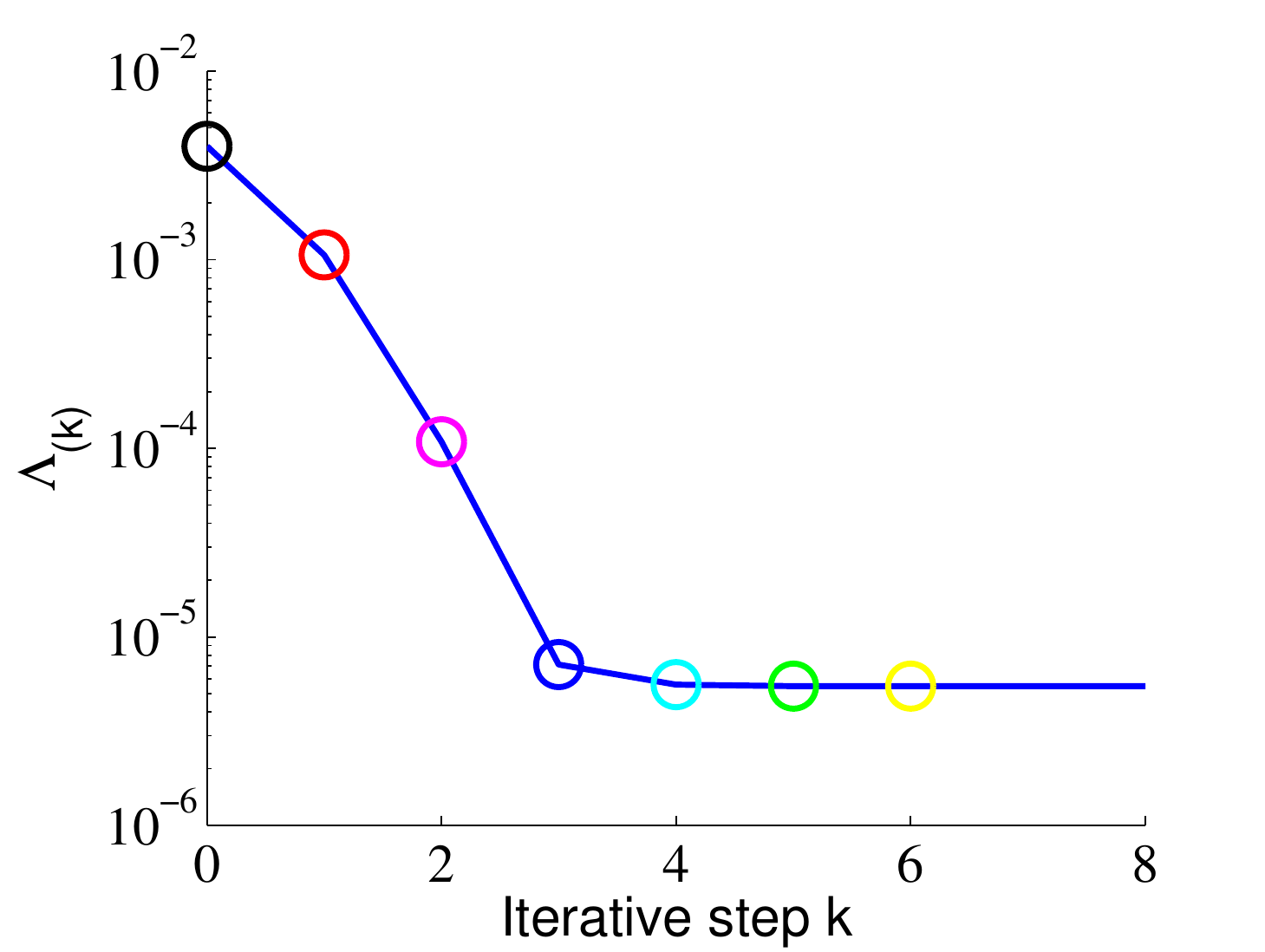}
(b)\includegraphics[scale=0.35]{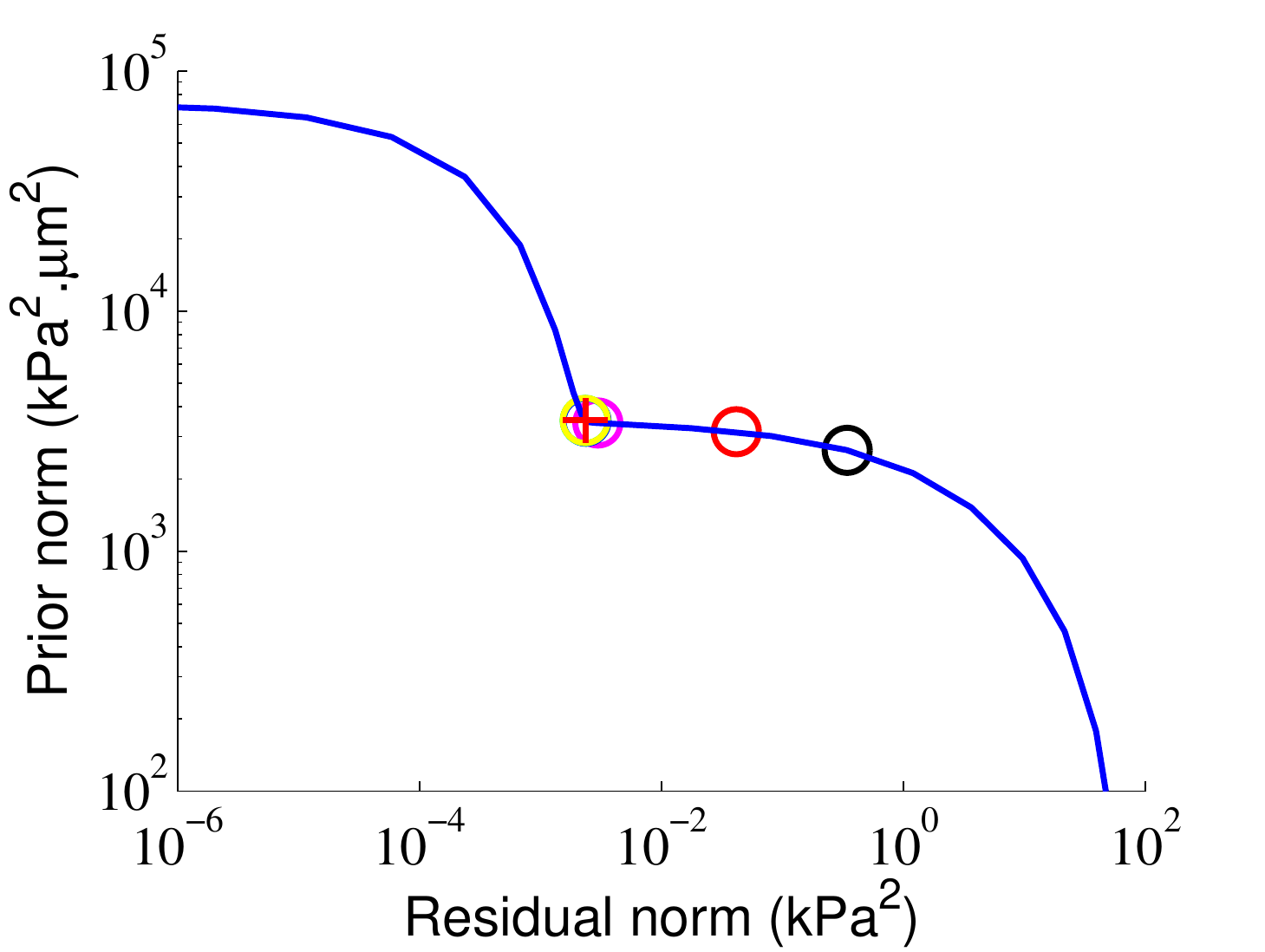}
(c)\includegraphics[scale=0.35]{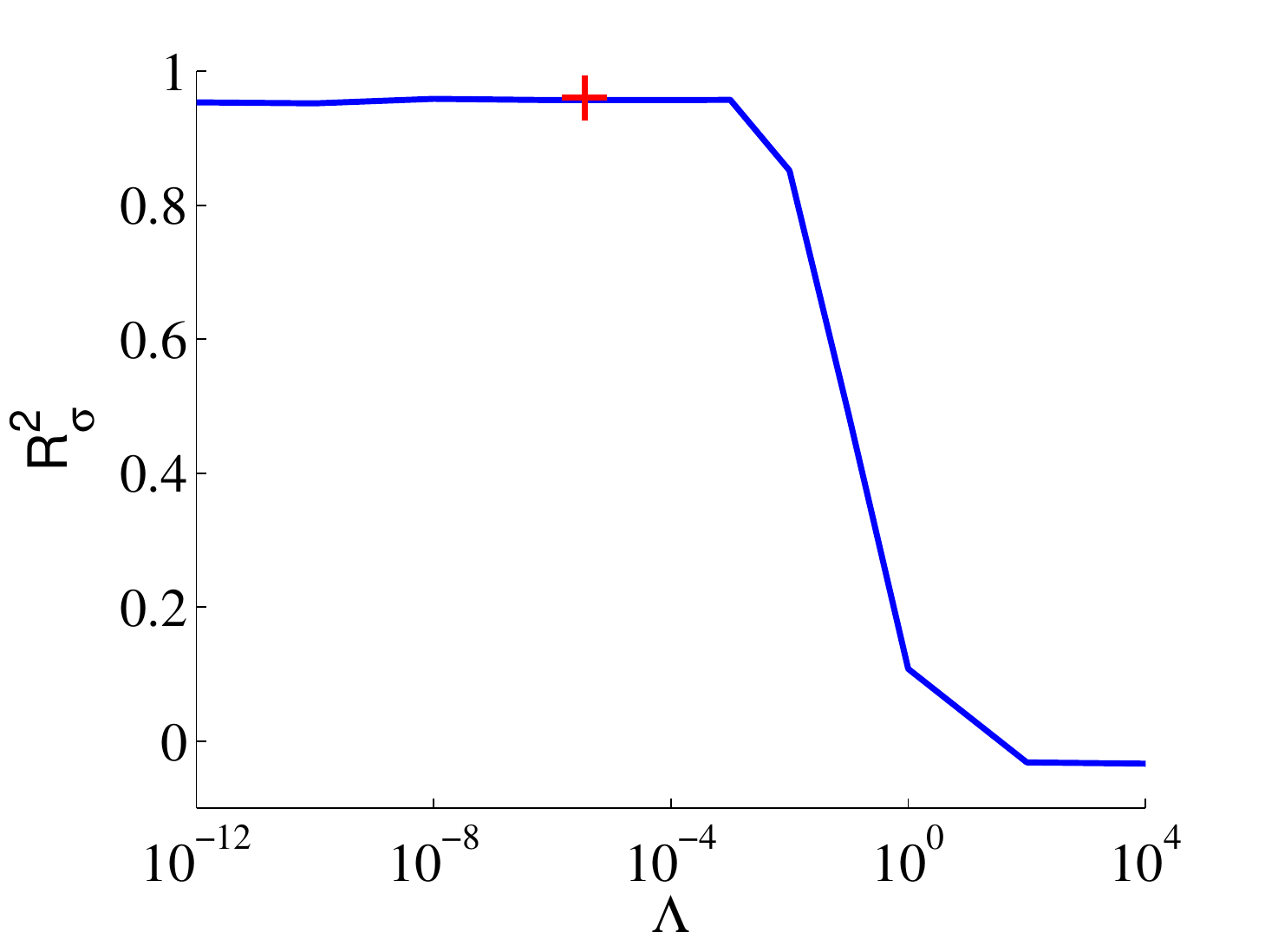}
}
\captionsetup{labelformat=suppfig}
\caption{\label{fig:opt:normal}
\textbf{Convergence: BISM.}
(a) $\Lambda_{(k)} = l^2 \,s^2_{(k)}/ s^2_{0\,(k)}$
\emph{vs.} iterative step $k$ in semilogarithmic scale.
(b) L-curve: effective prior norm
$\vec{{\sig}}_{\Pi}^t \, B^{-1} \, \vec{{\sig}}_{\Pi}$
\emph{vs.} residual norm $\| \vec{T} - A \vec{\sig}_{\Pi} \|^2$
in logarithmic scale. The colored circles correspond to the successive steps
depicted in (a). Black circle: initial condition; crossed circle:
asymptotic value, step $k=6$. The asymptotic value reached by the 
iterative method sits at the apex of the L-curve.
(c) Coefficient of determination $R^2_{\sig}$ \emph{vs.} $\Lambda$
in semilogarithmic scale. Red cross: asymptotic value
$\Lambda_{(\infty)}=5.5\, 10^{-6}$.
}
\end{adjustwidth}
\end{figure}

In \emph{Validation: numerical data}, we infer the stress field
of a viscous tissue with BISM. 
For initial values ${s^2}_{(0)}=10^{-1} \, \mathrm{kPa^2}$ and 
${s_0^2}_{(0)}=10^2 \, \mathrm{kPa^2 \,\mu m^2}$, the 
iterative resolution method converges in a few steps towards an 
optimal value of the regularization parameter $\Lambda_{(\infty)}=5.5\, 10^{-6}$ 
(${s^2}_{(\infty)}=4.7\, 10^{-7} \, \mathrm{kPa^2}$,
${s_0^2}_{(\infty)}=3.4\, 10^{-1} \, \mathrm{kPa^2 \,\mu m^2}$),
see Fig.~S\ref{fig:opt:normal}a

\begin{table*}
   \centering
\begin{adjustwidth}{-0.6in}{0in} 
\begin{tabular}{ccccccccccc}
  \hline
      & BISM & \emph{P1} & \emph{P2} & \emph{P3} & \emph{P4} & \emph{P5}
& \emph{P6} & \emph{S1} & \emph{S2} & \emph{S3} \\       
  \hline
    $\, \Lambda_{(\infty)} \,$  
& $\, 5.5\, 10^{-6} \,$ 
& $\, 8.1\, 10^{-4} \,$
& $\, 1.2\, 10^{-4} \,$ 
& $\, 3.7\, 10^{-6} \,$ 
& $\, 1.1\, 10^{-5} \,$ 
& $\, 8.8\, 10^{-6} \,$ 
& $\, 2.3\, 10^{-3} \,$ 
& $\, 5.6\, 10^{-4}  \,$ 
& $\, 1.2\, 10^{-5} \,$ 
& $\, 2.4\, 10^{-3} \,$\\ 
  \hline
\end{tabular}
\\
\bigskip
\begin{tabular}{ccccccccccc}
  \hline
      & BISM & \emph{N1}$_-$ & \emph{N1}$_+$ & \emph{N2} & \emph{N3} 
& \emph{N4} & \emph{N5} & \emph{N6}    \\       
  \hline
    $\, \Lambda_{(\infty)} \,$  
& $\, 5.5\, 10^{-6} \,$ 
& $\, 1.7\, 10^{-6} \,$
& $\, 5.6\, 10^{-6} \,$
& $\, 3.9\, 10^{-6} \,$ 
& $\, 1.9\, 10^{-3} \,$ 
& $\, 6.2\, 10^{-3} \,$ 
& $\, 2.6\, 10^{-5} \,$ 
& $\, 5.6\, 10^{-5} \,$  \\ 
  \hline
\end{tabular}
\vspace*{0.1cm}
\captionsetup{labelformat=supptable}
\caption{
\label{tab:Lambda} 
\textbf{Optimal values $\Lambda_{(\infty)} = l^2 \,s^2_{(\infty)}/ s^2_{0\,(\infty)}$}
of the regularization parameter for all statistical models and numerical
simulations studied in \emph{Results}.
}
\end{adjustwidth}
\end{table*}

When performing Tikhonov regularization, the optimal regularization
parameter is located at the apex of the L-curve \cite{hansen1992analysis}.
Here, the L-curve is a parametric plot of the prior norm 
$\vec{{\sig}}_{\Pi}^t \, B^{-1} \, \vec{{\sig}}_{\Pi}$ 
\emph{vs.} the residual norm $\| \vec{T} - A \vec{\sig}_{\Pi} \|^2$
as $\Lambda$ varies, see Fig.~S\ref{fig:opt:normal}b.
Superposing on the L-curve the points 
obtained by iteration, we find excellent agreement between the two methods. 
MAP estimation is more economical computationally since only
a few points are needed to determine the optimal parameters.
Successive iterations displace the representative circles towards 
smaller values of the residual norm along the L-curve. In order to obtain
convergence, we need to start from initial conditions located
on the right hand side of the apex. The representative circle runs away 
towards unphysically small values of $\Lambda$ when starting from
initial conditions located on the left hand side of the apex.
Another advantage of MAP estimation is to determine self-consistently the
values of $s^2$ and $s_0^2$, needed to compute $S_{\Pi}$ and the 
error bars $\delta \vec{\hat{\sig}}$.

\begin{figure}[!t]
\centering 
\begin{adjustwidth}{-0.6in}{-0.65in} 
\showfigures{
(a)\includegraphics[scale=0.35]{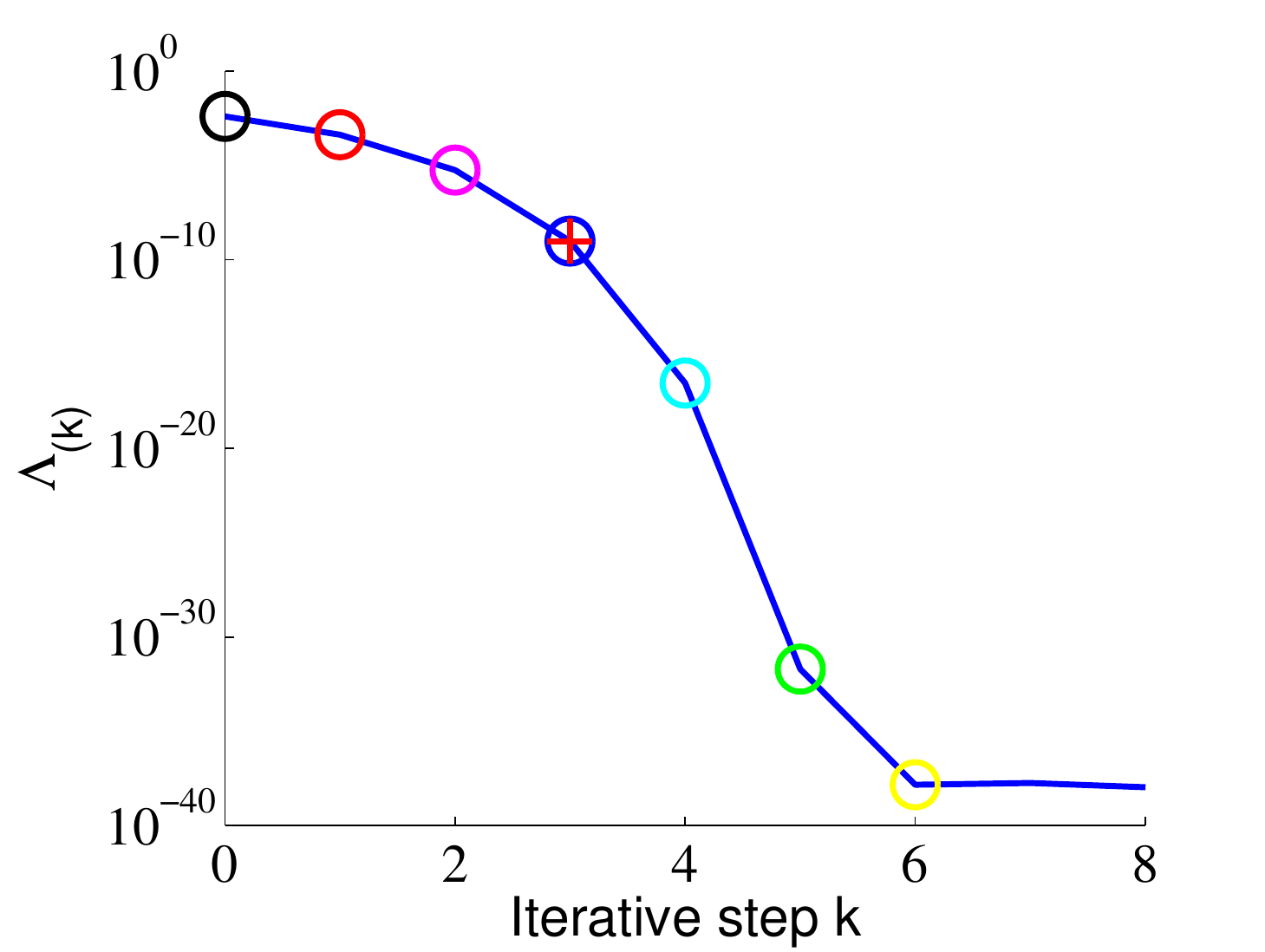}
(b)\includegraphics[scale=0.35]{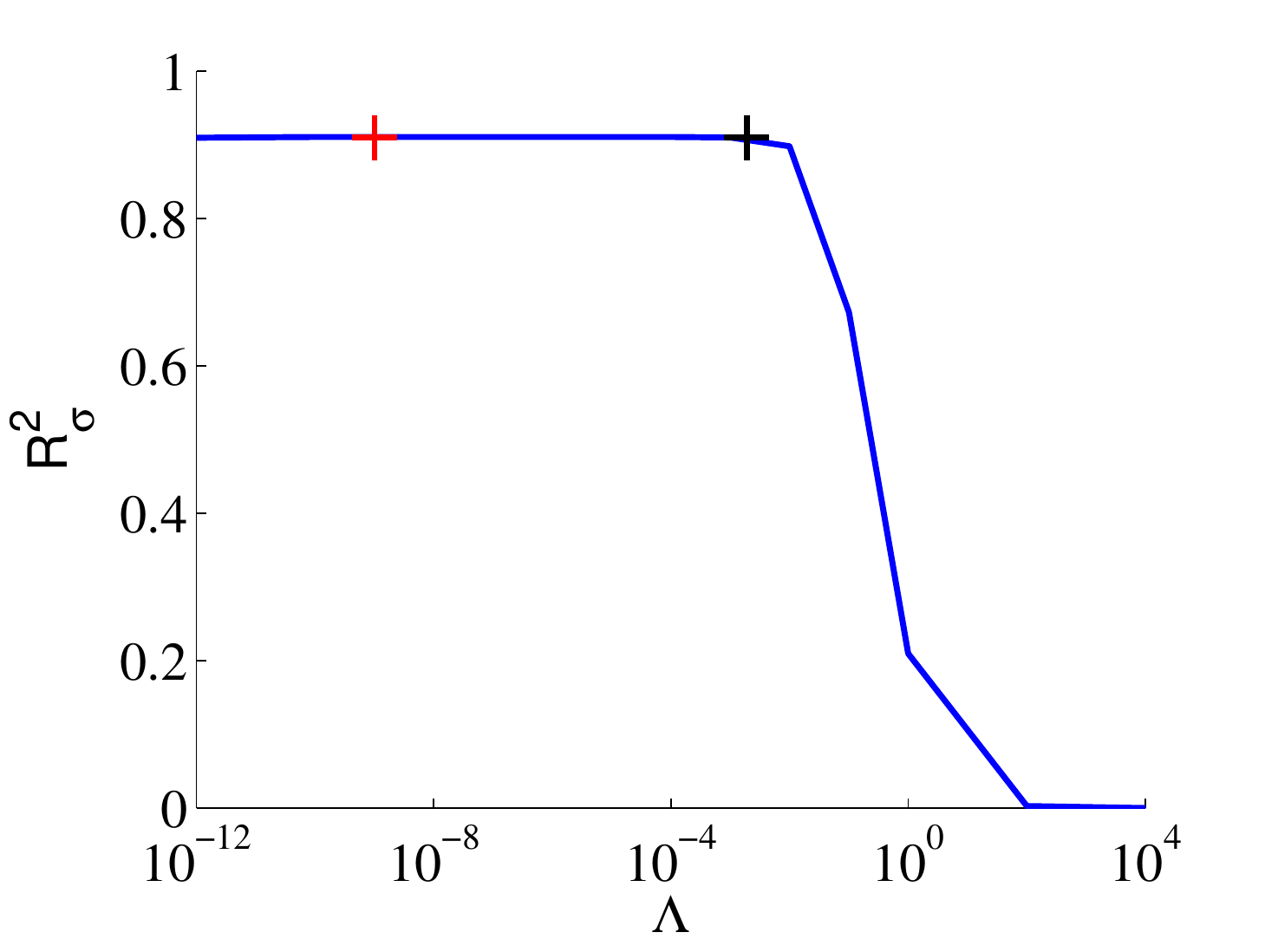}
(c)\includegraphics[scale=0.35]{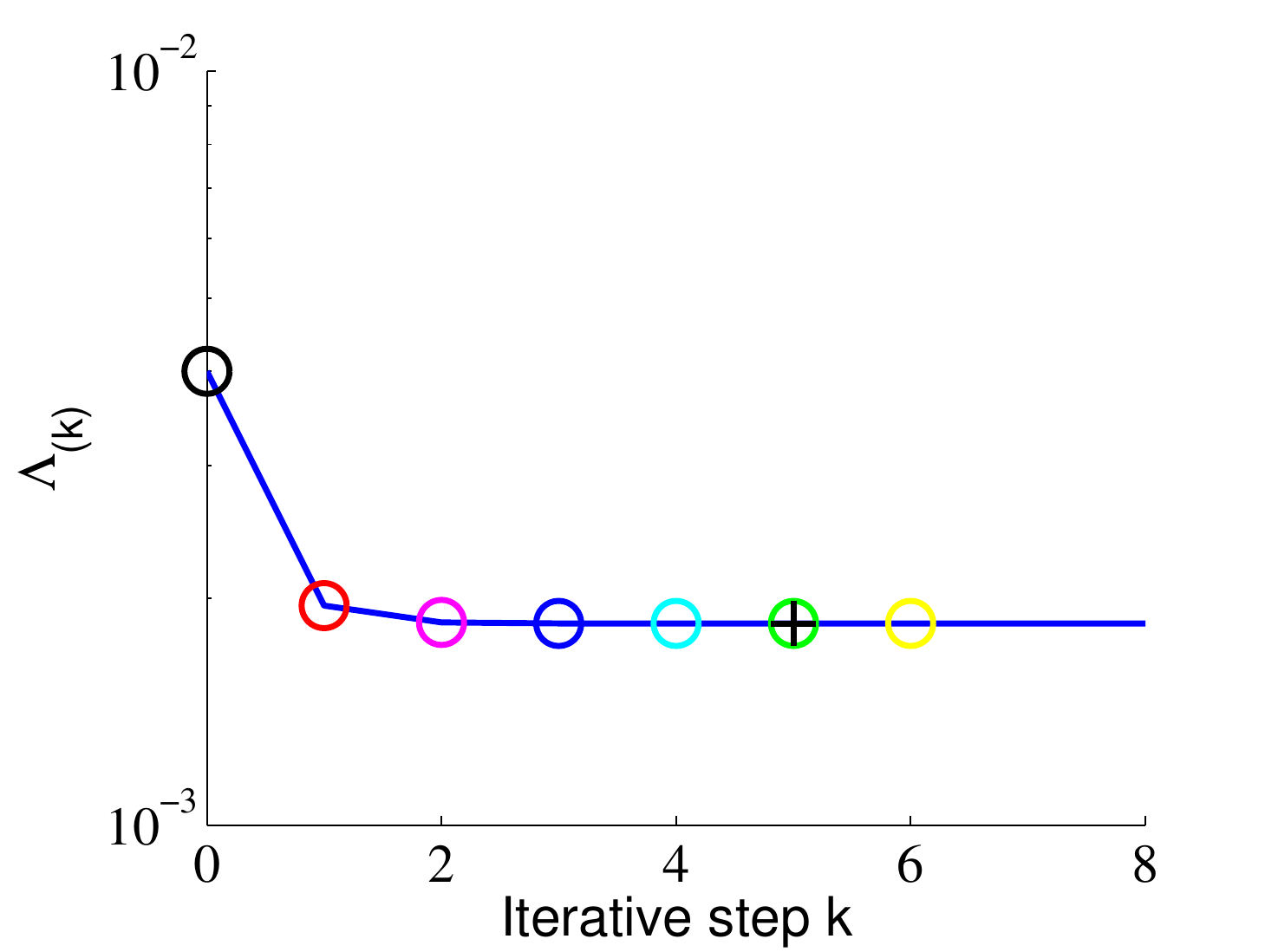}
}
\captionsetup{labelformat=suppfig}
\caption{\label{fig:opt:nonCV}
\textbf{Convergence: \emph{N3}.}
(a) $\Lambda_{(k)} = l^2 \,s^2_{(k)}/ s^2_{0\,(k)}$ \emph{vs.} iterative step $k$ 
in semilogarithmic scale. 
Red cross:  $k = 3$, $\Lambda = 1.0\, 10^{-9}$,  see text.
(b) Coefficient of determination $R^2_{\sig}$ \emph{vs.} $\Lambda$.
The red and black crosses locate the asymptotic values found in (a) and (c). 
(c) $\Lambda_{(k)}$ \emph{vs.} iterative step $k$ in semilogarithmic scale 
at constant $s^2 = 10^{-3}\, \mathrm{kPa}^2$, with 
$\Lambda_{(\infty)}=1.9\, 10^{-3}$.
The black and colored circles in (a) and (c) correspond respectively
to the initial condition and to successive iterative steps 
}
\end{adjustwidth}
\end{figure}

The value $\Lambda_{(\infty)}=5.5\, 10^{-6} \ll 1$ gives more importance to the
likelihood than to the prior. Inferring the stress field at fixed values
of $s_0^2$ and $s^2$, we plot the coefficient of determination $R_{\sigma}^2$
\emph{vs.} $\Lambda$ (Fig.~S\ref{fig:opt:normal}c) 
and find that the accuracy of inference is excellent as
soon as $\Lambda < 10^{-3}$, and that further decreasing $\Lambda$ 
no longer improves accuracy. BISM is robust against
variations of the ratio $s^2/s^2_0$ provided that the weight 
given to information contained in the prior is small enough.

This pattern describes faithfully the convergence of the resolution method
for most of the statistical models and numerical data sets we investigated:
convergence occurs within a few steps towards optimal values of the 
regularization parameter in the range $[10^{-6}, 10^{-3}]$, 
see Table~S\ref{tab:Lambda}.

Variations \emph{P6}, \emph{N3} and \emph{N4}, which
correspond to changes related to boundary conditions, differ
in the following way. The resolution method converges fast, but to an 
unphysically small value of $\Lambda_{(\infty)}$
(of the order of $10^{-38}$ for \emph{N3}, see Fig.~S\ref{fig:opt:nonCV}a).
Since the coefficient of determination  still plateaus close to $1$
when $\Lambda < 10^{-2}$ (Fig.~S\ref{fig:opt:nonCV}b),
selecting an early step (typically $k = 3$) for inference
works well, and gives the value $\Lambda_{(\infty)} = 1.0\, 10^{-9}$.

However, setting $s^2$ to a constant value and ignoring 
Eq.~(19) allows to self-consistently determine 
$s_0^2$ iteratively using Eq.~(20) only.
When dealing with experimental data, a natural estimate of
$s$ is the experimental error bar on traction force data.
For the numerical data sets \emph{P6}, \emph{N3} and \emph{N4},
we use $s^2 = 10^{-3}\, \mathrm{kPa}^2 \approx s^2_{\mathrm{exp}}$, 
the noise amplitude (see Table~S\ref{tab:Lambda}).
For \emph{N3}, iterations over $\vec{\sig}_{\Pi\,(k)}$ and
$s^2_{0\,(k)}$ exhibit rapid convergence towards
$\Lambda_{(\infty)}=1.9\, 10^{-3}$ (see Fig.~S\ref{fig:opt:nonCV}c),
on the plateau of the $R_{\sig}^2(\Lambda)$ curve. 
In this case, error bars $\delta {\hat{\sig}}$ on the inferred stress 
are less meaningful since they depend on the numerical value used for $s^2$.
Of course if $s^2$ is fixed by the experiment then error bars 
are more reliable. Due to the large plateau of the $R_{\sig}^2(\Lambda)$ curve,
both parameter determination methods yield the same results.

\begin{figure}[!t]
\showfigures{
(a)\includegraphics[scale=0.3]{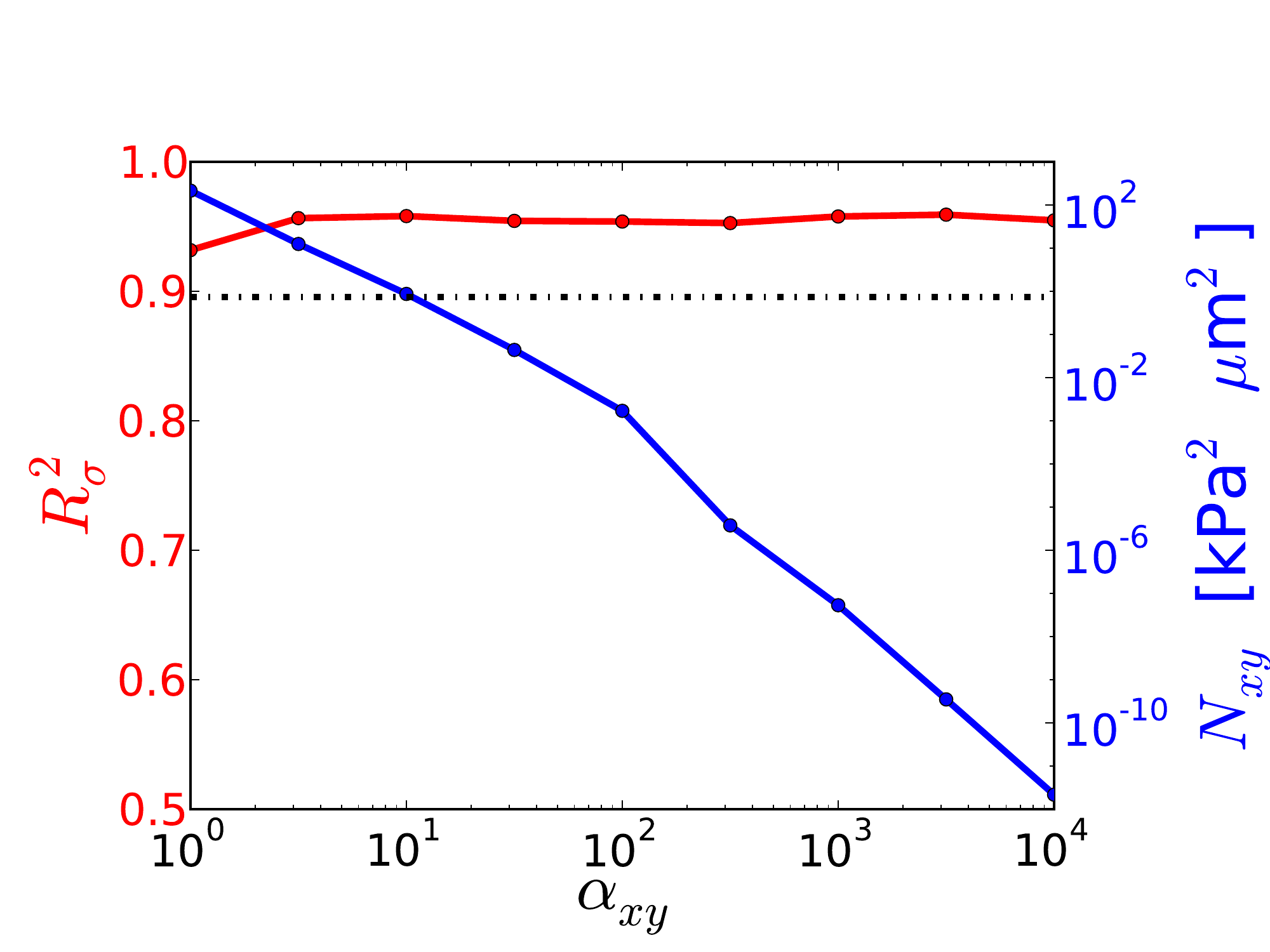}
\hspace*{1cm}
(b)\includegraphics[scale=0.3]{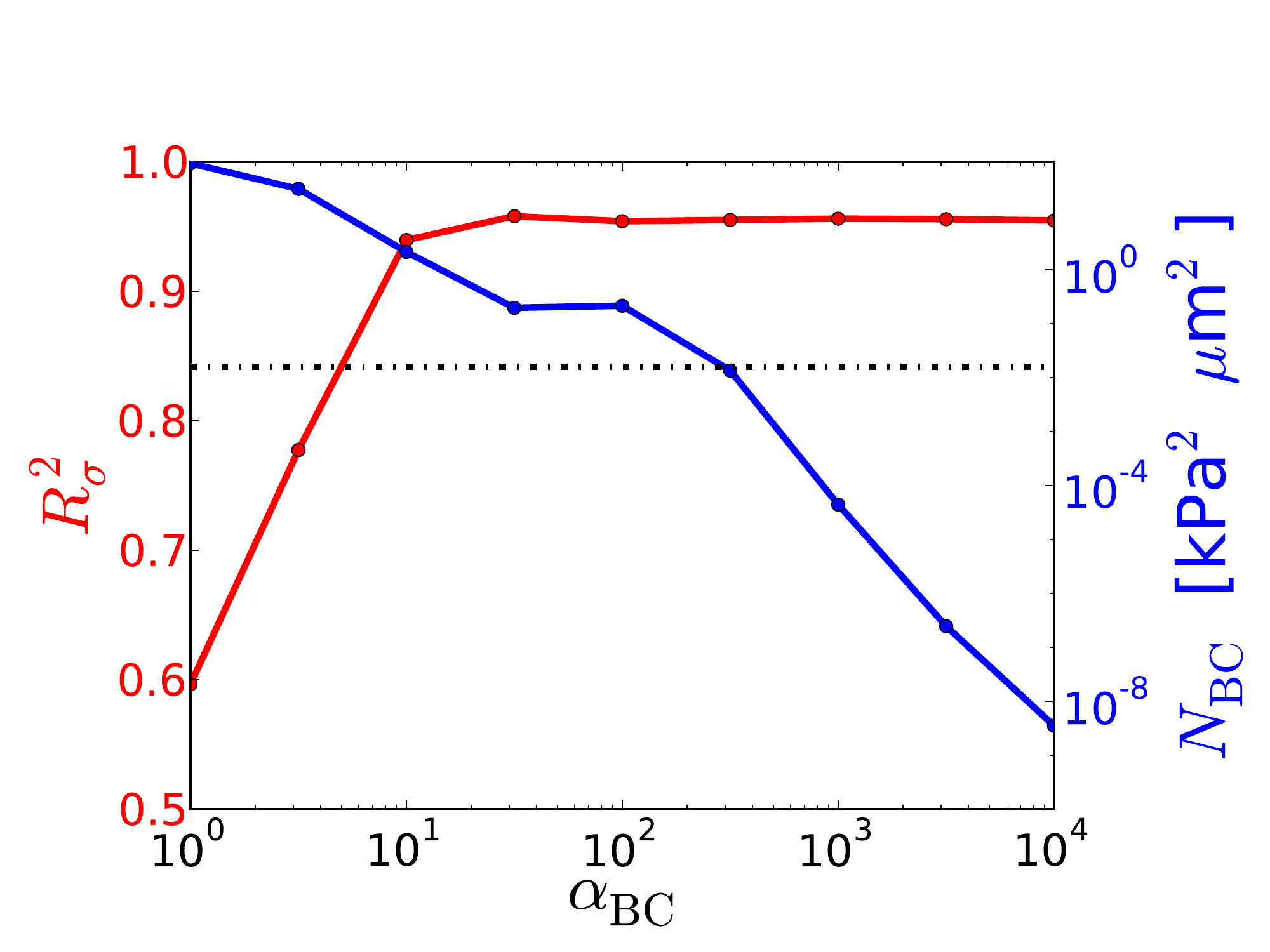}
}
\captionsetup{labelformat=suppfig}
\caption{
\textbf{Hyperparameters $\axy$ and $\aBC$: BISM.}
(a) $R^2_{\sig}$ (red) and $N_{xy}$ (blue) \emph{vs.} $\axy$.
(b) $R^2_{\sig}$ (red) and $N_{\mathrm{BC}}$ (blue) \emph{vs.} $\aBC$. 
On both graphs, the black dashed line gives the reference values, 
$N_{xy}^{\mathrm{num}} = 7.4\;10^{-1} \, \mathrm{kPa^2 \,\mu m^2}$ and 
$N_{\mathrm{BC}}^{\mathrm{num}} = 1.6\;10^{-2} \, \mathrm{kPa^2 \,\mu m^2}$ 
computed for the numerical data set.
}
\label{fig:opt:alpha:beta}
\end{figure} 

\subsection{Hyperparameters $\axy$ and $\aBC$}
\label{sec:opt:alpha:beta}

In principle the values of  $\axy$ and $\aBC$ could also be determined
self-consistently. For simplicity, we set these hyperparameters to constant,
large values.

In order to determine which hyperparameter values  will
ensure that stress tensor symmetry and boundary conditions are respected,
we perform BISM on the \emph{Viscous} data set for a large range $[1, 10^4]$
of fixed values of $\axy$ and $\aBC$. In addition to $R_{\sig}^2$, 
we define the partial prior norms
$N_{xy} = \| \vec{\sigma}_{xy} -\vec{\sigma}_{yx} \|^2$  
and $N_{\mathrm{BC}} = \| \vec{\sig}_{BC} \|^2$ as specific
measures of accuracy.  
Fig.~S\ref{fig:opt:alpha:beta}a shows how $R_{\sig}^2$ and $N_{xy}$ 
vary as a function of $\axy$. The partial prior norm 
$N_{xy}$ decreases with $\axy$. Above $\axy \simeq 10$,  $N_{xy}$
becomes smaller than the reference value $N_{xy}^{\mathrm{num}}$
computed on the (noisy) data set. 
Since the global measure $R^2_{\sig}$ plateaus for $\axy > 10^1$,
values of $\axy$ larger than $10^1$ ensure equality of the
shear stress components. Note that as $\axy$ decreases, $R^2_{\sig}$ decreases 
towards the limit $R^2_{\sig}(\axy = 0) = 0.75$ found for variation \emph{P5}.
The same reasoning applies to $\aBC$, for which $R^2_{\sig}$ plateaus for 
$\axy > 3 \, 10^1$, and the reference value
$N_{\mathrm{BC}}^{\mathrm{num}}$  computed on the noisy data set is 
crossed above $300$, see Fig.~S\ref{fig:opt:alpha:beta}b. 

In practice, we use the conservative values $\axy = 10^3$ and $\aBC = 10^3$
for all statistical models and data sets.

\FloatBarrier
\section{Supplementary figures}
\label{sec:supp:fig}

\begin{figure}[h!]
\centering 
\showfigures{
\includegraphics[scale=0.47]{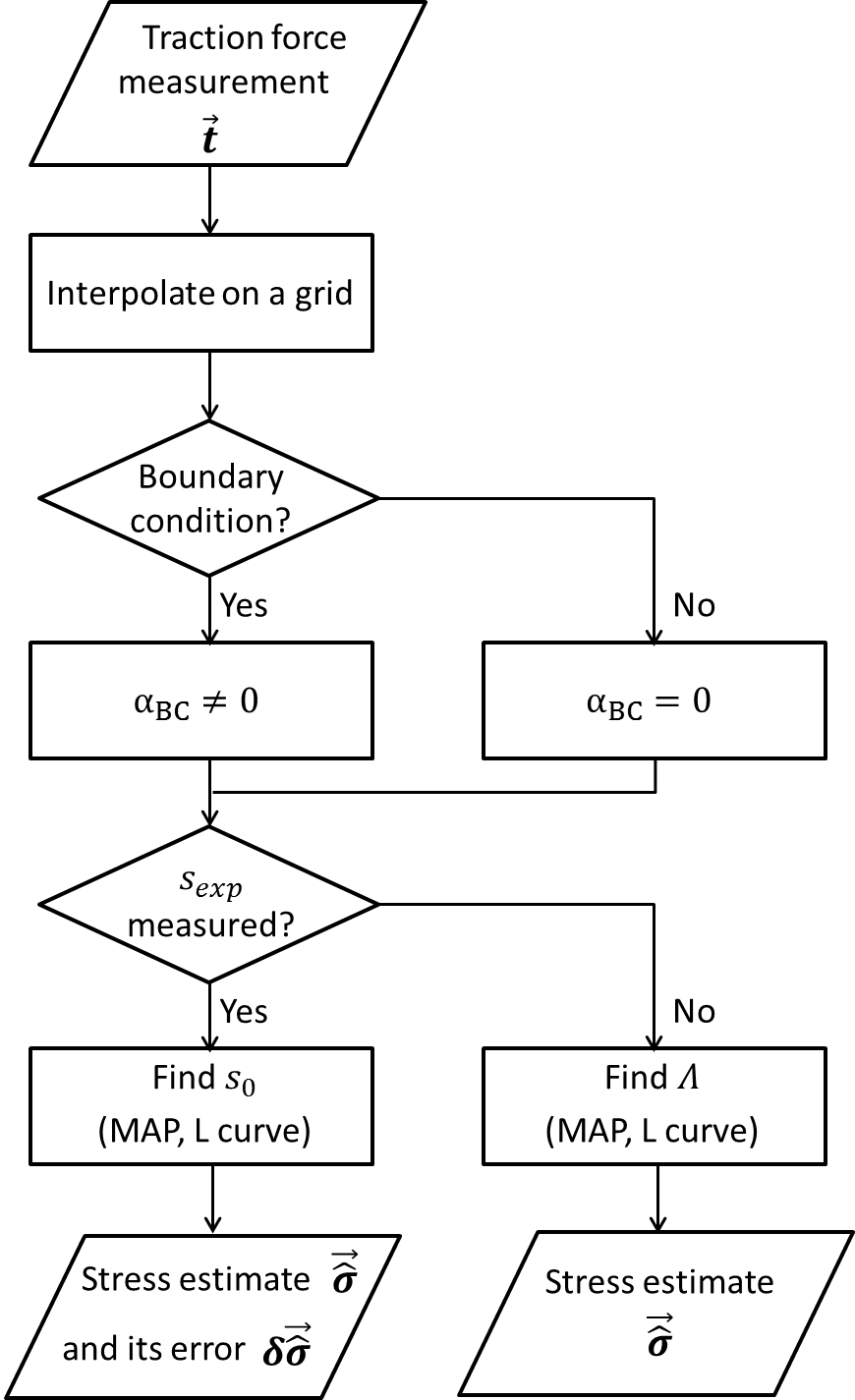}
}
\captionsetup{labelformat=suppfig}
\caption{\label{fig:flowchart}
\textbf{Flowchart of the algorithm.}
}
\end{figure}

\begin{figure}[!t]
\centering 
\showfigures{
\includegraphics[scale=0.5]{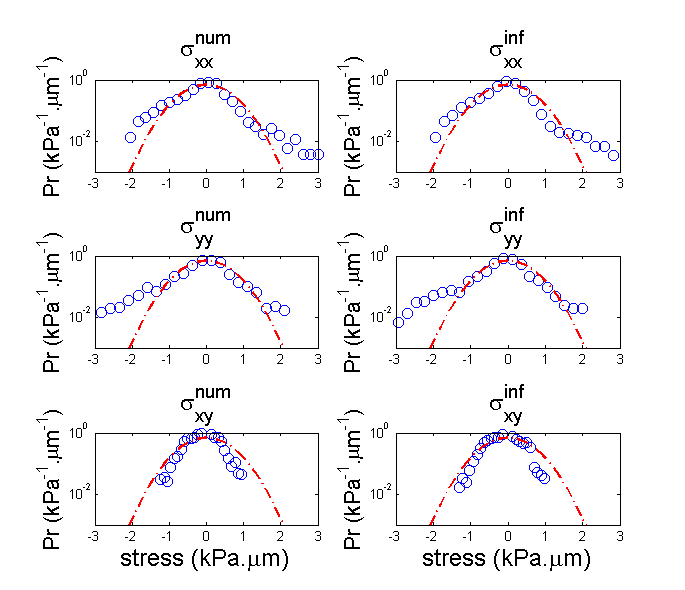}
}
\captionsetup{labelformat=suppfig}
\caption{\label{fig:hist}
\textbf{Stress probability distribution functions.} 
We compare the distributions  (blue circles) of simulated (left column) and 
inferred (right column) stresses, for each component,
in the \textit{Viscous} case. The red dashed line corresponds to the 
Gaussian prior distribution function 
with a standard deviation \mbox{$s_0=0.57 \, \mathrm{kPa.\mu m}$}.
}
\end{figure}

\begin{figure}[!b]
\centering 
\showfigures{
\includegraphics[scale=0.4]{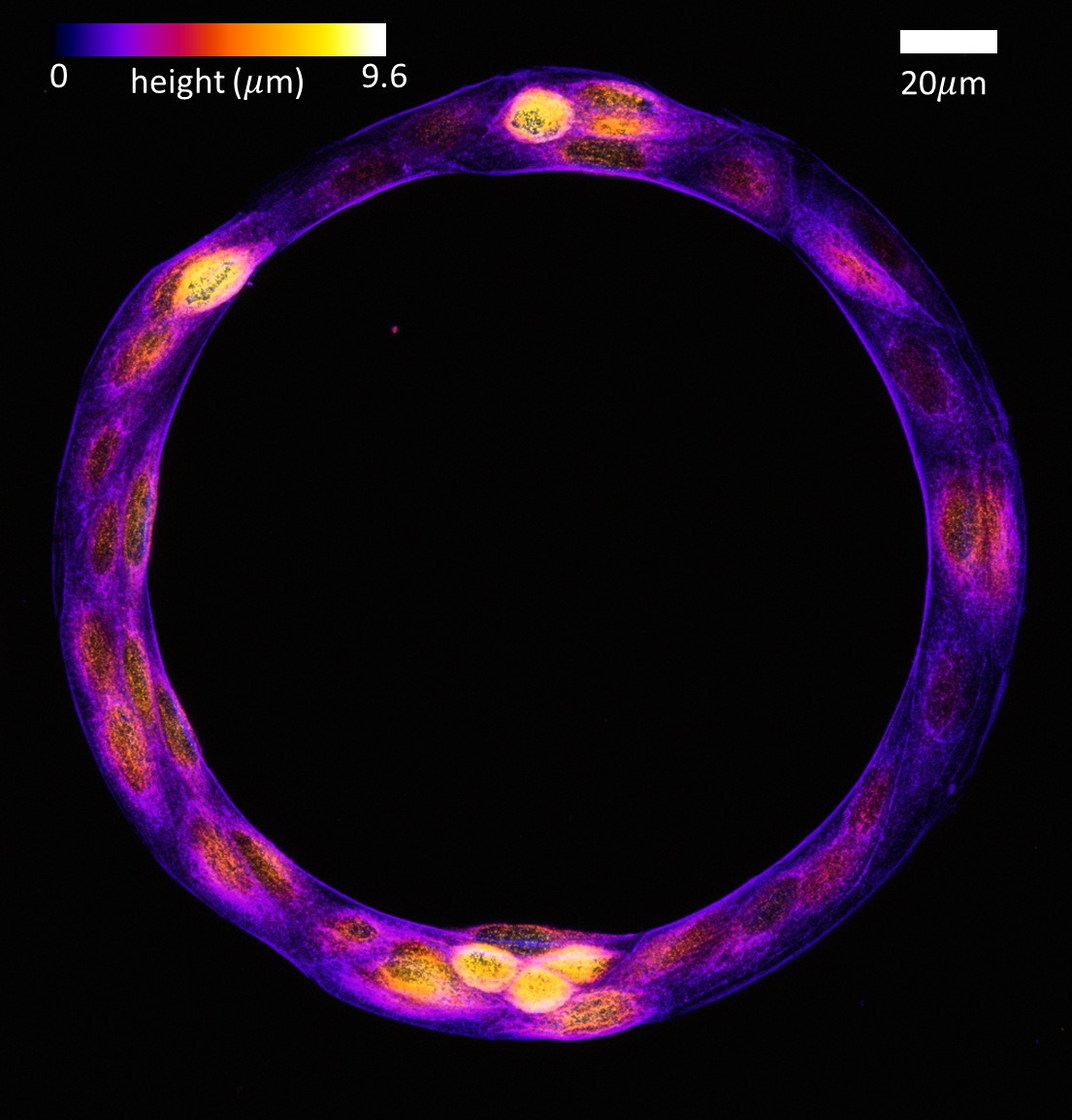}
}
\captionsetup{labelformat=suppfig}
\caption{\label{fig:height}
\textbf{Height variation of the annular cell monolayer.} Z-stack measurement: 
each color corresponds to a height bin in the z direction with a resolution 
of $0.4 \,\mathrm{\mu m}$. 
Scale bar: $20\,\mathrm{\mu m}$.
}
\end{figure}

\begin{figure*}[!t]
\centering 
\begin{adjustwidth}{-0.6in}{-0.65in} 
\showfigures{
(a)\includegraphics[scale=0.6]{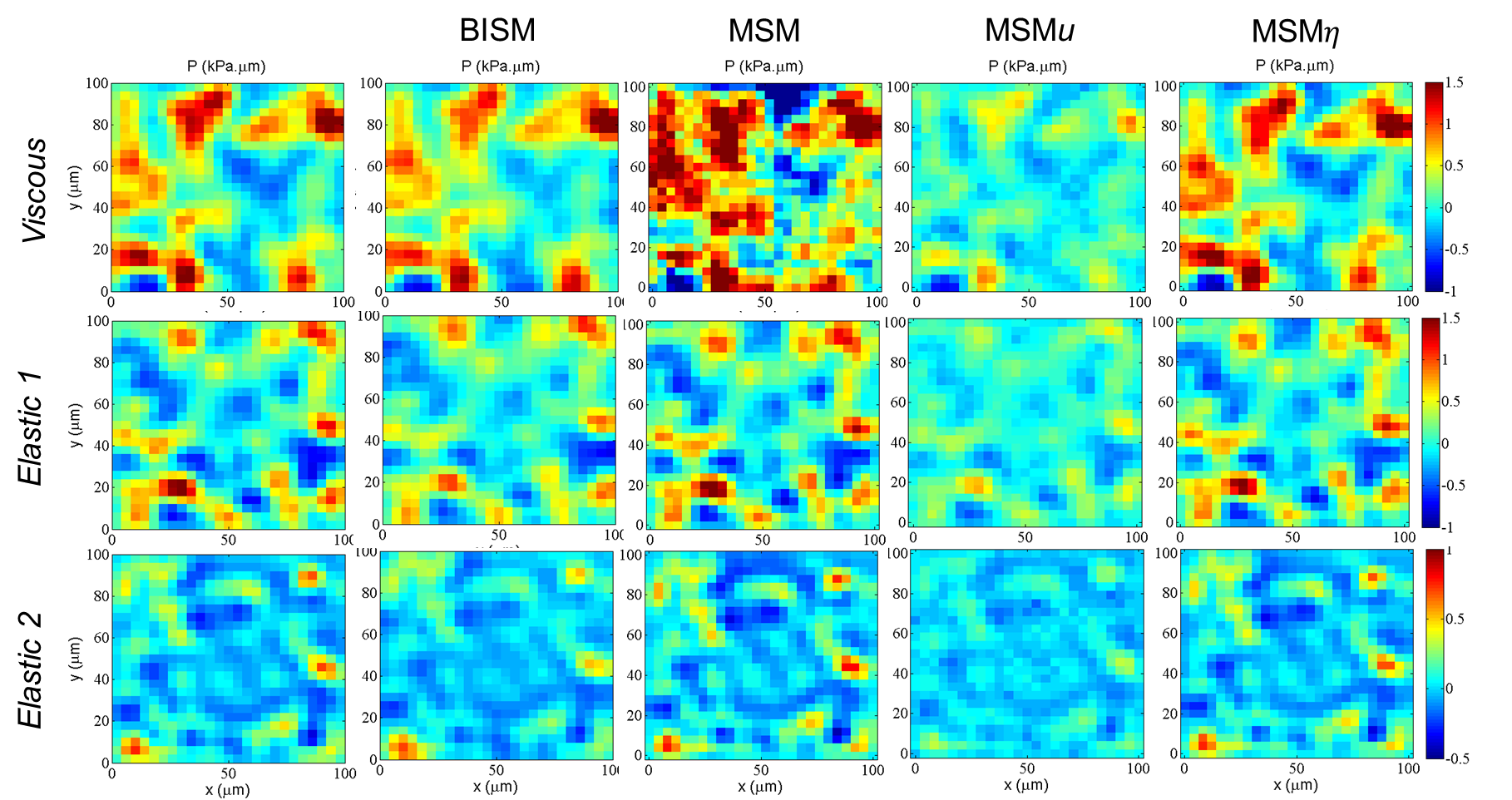}
(b)\includegraphics[scale=0.6]{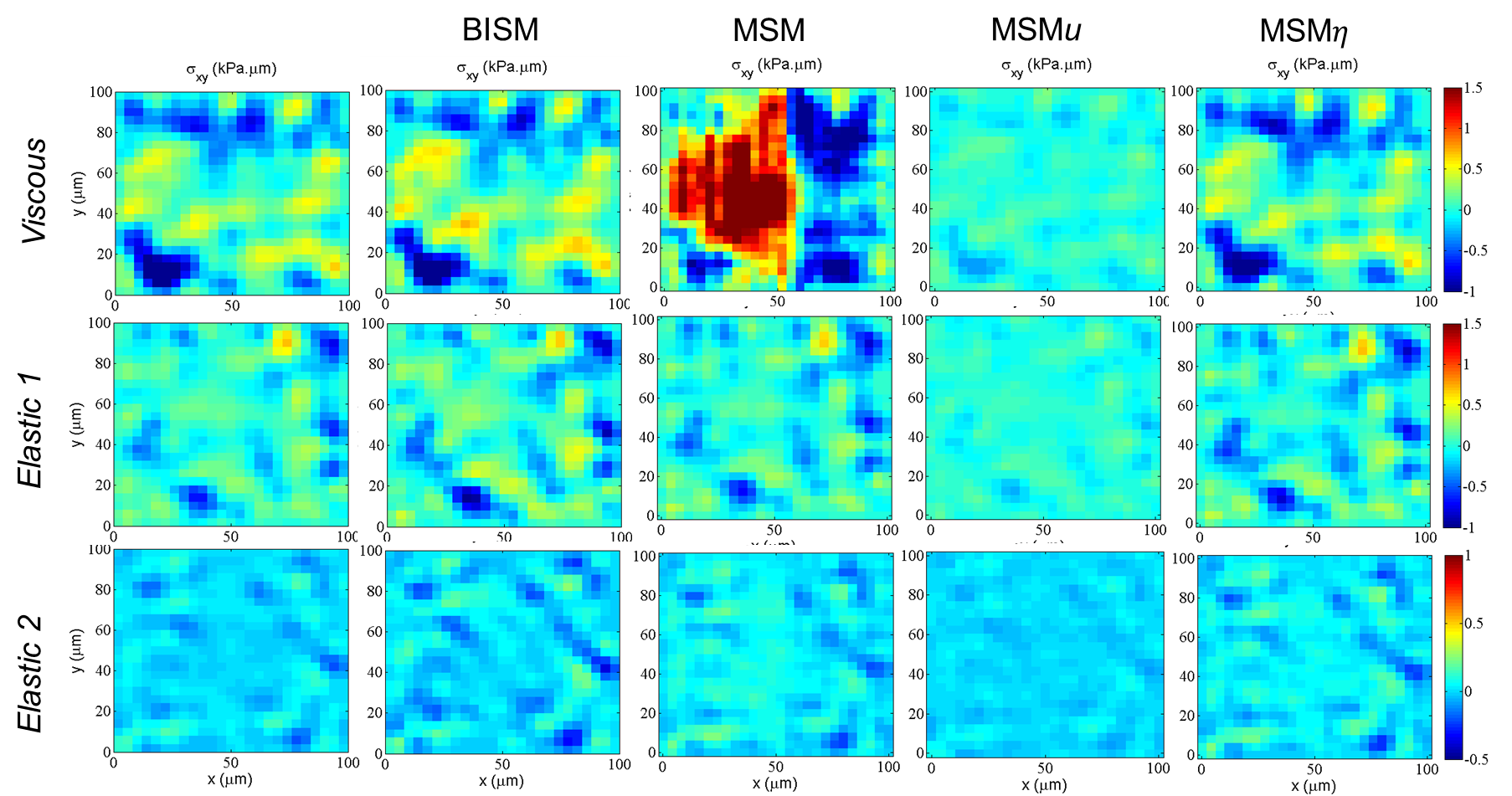}
}
\end{adjustwidth}
\end{figure*}

\begin{figure}[!t]
\centering 
\begin{adjustwidth}{-1.0in}{-0.65in} 
\showfigures{
(c)\includegraphics[scale=0.6]{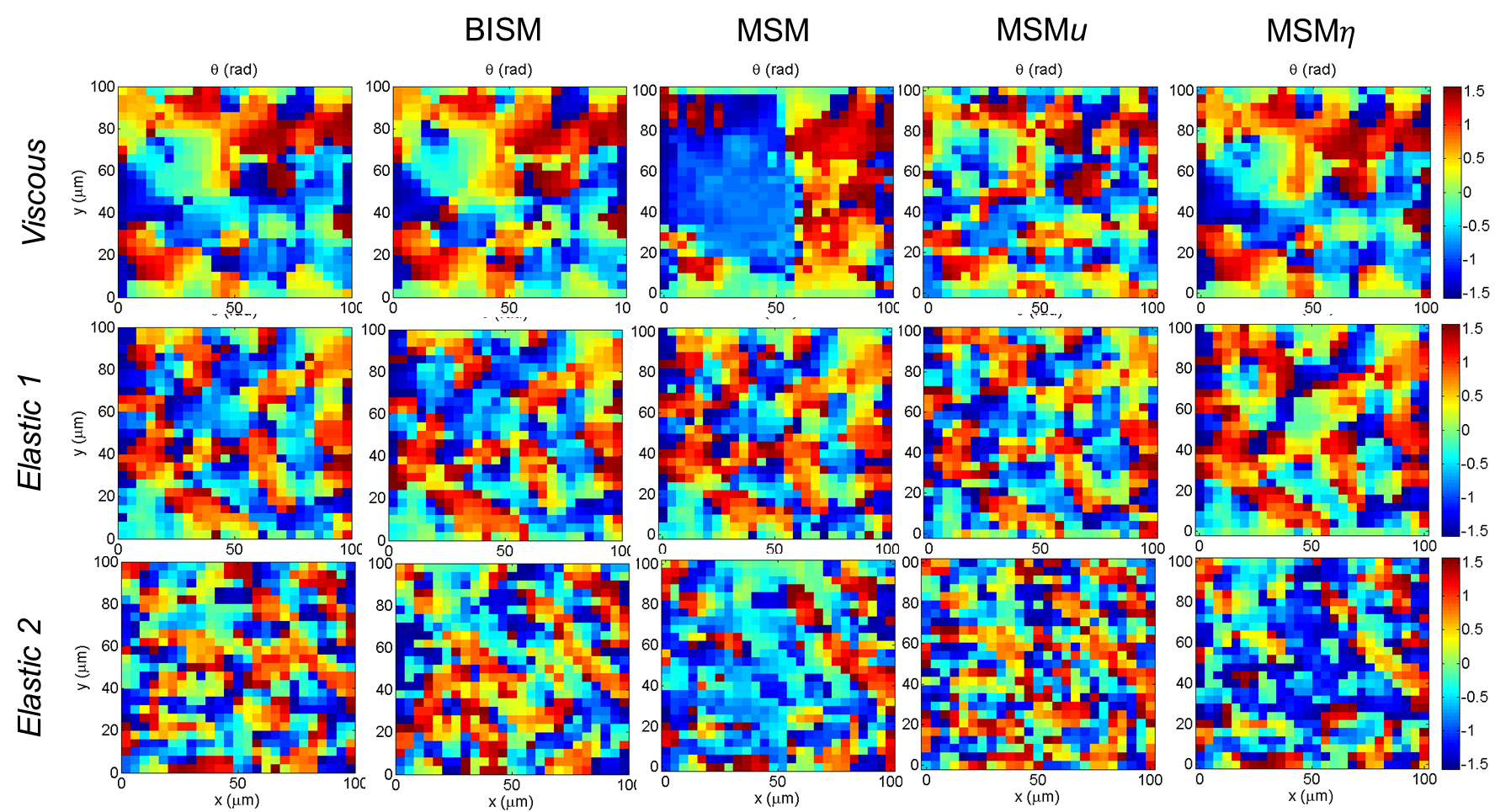}
}
\captionsetup{labelformat=suppfig}
\caption{\label{fig:disc}
\textbf{Comparison between BISM and monolayer stress microscopies.}
Numerical data (first column) for: (a) pressure; (b) shear stress; 
(c) stress orientation (angle $\theta$ between the principal stress eigenvector
and the $x$ axis, $\theta \in [- \pi/2, \pi/2]$) 
are compared with inferred data obtained with BISM, 
MSM, MSM$u$ and MSM$\eta$ (see text for definitions,
spatial resolution: $4\,\mu$m).
Traction force data sets were obtained with material parameter values
$\eta=10^{3} \, \mathrm{kPa \, \mu m\, s}$, $\eta'=\eta$ (\emph{Viscous});
$E=10^2 \, \mathrm{kPa \,\mu m}$, $\nu_{2 \mathrm{D}}=0.5$ \cite{Tambe2011}
(\emph{Elastic 1}), $E=10 \, \mathrm{kPa \,\mu m}$,
$\nu_{2 \mathrm{D}}=0.5$ \cite{Moussus2014} (\emph{Elastic 2}).
A white noise of  relative amplitude $5\,\%$ is added in all cases.
}
\end{adjustwidth}
\end{figure}


\end{document}